\documentclass[
  aps,
  prd,
  11pt,
  superscriptaddress,
  notitlepage,
  onecolumn,
  amsmath,
  nofootinbib
]{revtex4-1}

\AtBeginDocument{
  \linespread{1.05}\selectfont
}

\usepackage[
  a4paper,
  left=2.3cm,
  right=2.3cm,
  top=2.5cm,
  bottom=2.5cm
]{geometry}

\usepackage{microtype} 
\usepackage{graphicx}
\usepackage{xcolor}
\colorlet{RED}{red}
\usepackage{orcidlink}
\usepackage{enumerate}
\usepackage{enumitem}
\usepackage{ulem}
\usepackage{amssymb} 
\usepackage[svgnames]{xcolor} 
\usepackage{tikz}
\usetikzlibrary{arrows.meta,positioning,fit,calc}

\definecolor{linkcolor}{rgb}{0.0,0.3,0.5}
\definecolor{linkcolor}{rgb}{0.0,0.3,0.5}
\definecolor{mypurple}{RGB}{143, 116, 210}

\graphicspath{{Figures/}}

\newif\ifshowchanges

\showchangestrue

\newcommand{\hias}{School of Fundamental Physics and Mathematical Sciences, Hangzhou Institute for Advanced Study, University of Chinese Academy of Sciences, Hangzhou 310024, China}

\begin{document}

\title{Massive boson stars: Waveform-based branch diagnosis with neural reconstruction}

\author{Bo-Xuan Ge
\orcidlink{0000-0003-0738-3473}}
\email{bo-xuan.ge@ucas.ac.cn}
\affiliation{\hias}

\begin{abstract}
We investigate whether the dynamical merger branch of massive boson-star
collisions can be inferred from the gravitational waveform without supplying
the intrinsic source parameters.  Using an existing equal-mass, head-on
numerical-relativity catalogue, we construct branch-conditioned reduced-basis
surrogates for the real $(2,0)$ and $(4,0)$ modes and perform
parameter-blind waveform inference.  On a frozen test set of 60 held-out
samples, the correct branch is recovered for all samples using either mode,
showing that the branch information is not confined to the dominant $(2,0)$
channel.  The inference also recovers $|\phi_c|$ with a mean relative error
of $4.30\%$ and selects the correct $\lambda$ from
$\{10,50,100,300\}$ in $68.3\%$ of the test samples.  These results
demonstrate waveform-based branch and source-parameter inference within the
present noiseless numerical-relativity catalogue.
\end{abstract}

\maketitle

\section{Introduction}
\label{sec:introduction}

The direct detection of gravitational waves has opened a new observational
window on strong-field gravity and compact-object dynamics
~\cite{Abbott:2016blz,LIGOScientific:2018mvr,LIGOScientific:2020ibl,
LIGOScientific:2021djp}.  These observations have made it possible to test
general relativity in the highly nonlinear regime, to probe the astrophysical
origin of compact binaries, and to ask whether all compact objects observed
through gravitational waves are necessarily black holes or neutron stars.  In
this broader context, exotic compact objects provide useful theoretical
laboratories for exploring possible alternatives to the standard compact-binary
picture.

Boson stars are among the simplest and most widely studied examples of such
objects.  They are self-gravitating configurations of scalar fields, originally
introduced in the context of mini boson stars
~\cite{Kaup:1968zz,Ruffini:1969qy}, and later generalized to include
self-interactions that can support astrophysically massive configurations
~\cite{Colpi:1986ye}.  Independently of their possible connection to dark
matter, boson stars are useful theoretical laboratories for horizonless
compact objects and black-hole mimickers.  Their isolated structure,
stability, and phenomenology have been investigated in a wide range of
models~\cite{Seidel:1993zk, Schunck:1999pm,
Sanchis-Gual:2019ljs, Siemonsen:2023hko, Colpi:1986ye, Seidel:1990jh,
Kobayashi:1994qi, Ryan:1996nk, Schunck:1996he, Balakrishna:1997ej,
Yoshida:1997qf, Schunck:1999zu, Schunck:2003kk, Balakrishna:2006ru,
Balakrishna:2007mr, Hartmann:2012da, Siemonsen:2020hcg,
Evstafyeva:2025mvx, Marks:2025jpt, Evstafyeva:2023kfg, Marks:2025xxv,
Marks:2025jit, Ma:2024olw, Ding:2023syj, Liang:2022mjo,
Zhang:2023qxf, Zhang:2025xnl, deSa:2025nsx, Herdeiro:2025lwf,
Herdeiro:2024myz, Ildefonso:2023qty, Brito:2023fwr}, while their
binary dynamics have revealed a rich variety of nonlinear
behaviours~\cite{Palenzuela:2006wp,
Palenzuela:2007dm, Palenzuela:2017kcg, Helfer:2021brt,
Sanchis-Gual:2020mzb, Bezares:2022obu,
Croft:2022bxq, Sanchis-Gual:2022zsr, Evstafyeva:2022bpr,
Siemonsen:2023age, Ge:2024fum, Evstafyeva:2024qvp,
Brito:2025rld, Jaramillo:2022zwg, Damour:2025oys, Ning:2026qxs,
Ge:2026wzh, Ge:2025btw,Ge:2026fki}.

A central lesson from numerical studies of boson-star binaries is that their
merger dynamics are not described by a single universal outcome.  Depending
on the compactness, self-interaction strength, scalar-field structure, and
initial configuration, a binary may form a long-lived non-black-hole remnant,
collapse to a black hole after the two scalar configurations have interacted,
or undergo collapse before the two stars come into contact.  In the
quartically self-interacting massive-boson-star catalogue considered in this
work, these possibilities define three merger-outcome branches.  We denote
the branch that leaves a boson-star remnant by $\mathrm{BS}_{\rm post}$, the
branch in which a black hole forms after contact by $\mathrm{BH}_{\rm post}$,
and the branch in which the two stars collapse to black holes before contact
by $\mathrm{BH}_{\rm pre}$.

These branches are defined by the nonlinear spacetime and matter dynamics.
In Ref.~\cite{Ge:2025btw}, the outcome of each simulation was identified
using numerical-relativity diagnostics of the spacetime and scalar field,
including collapse behaviour, black-hole formation, and the nature of the
late-time remnant.  Such diagnostics determine the branch once the full
nonlinear evolution is available.  They do not, however, directly answer a
waveform-level inverse question: whether the emitted gravitational radiation
alone contains enough information to identify the dynamical route followed by
the merger.

This question is particularly important because the three branches are not
merely different labels assigned to otherwise similar waveforms.  The
$\mathrm{BS}_{\rm post}$ branch describes a merger that remains horizonless
during the simulated time interval, whereas both $\mathrm{BH}_{\rm post}$ and
$\mathrm{BH}_{\rm pre}$ end in black-hole formation.  The latter two branches
are especially nontrivial to distinguish, since they share the same broad
late-time fate but differ in whether collapse occurs after scalar-field
contact or before contact.  If their waveforms can be separated, then the
gravitational radiation retains information not only about the final compact
object, but also about the dynamical pathway through which that final state
was reached.

The aim of the present paper is therefore not to construct a production-ready
waveform template for boson-star binaries.
Instead, we ask whether the gravitational-wave signal can be used to
identify the underlying merger-outcome branch when the intrinsic source
parameters are not supplied a priori.  We retain a reconstruction-based
strategy: a branch-conditioned surrogate generates candidate waveforms, and
the preferred branch and source parameters are inferred from waveform
agreement rather than from a direct classification in the catalogue-parameter
space.

The numerical-relativity data are taken from
Ref.~\cite{Ge:2025btw}, which provides head-on equal-mass mergers of
quartically self-interacting massive boson stars together with the
$\mathrm{BS}_{\rm post}$, $\mathrm{BH}_{\rm post}$, and
$\mathrm{BH}_{\rm pre}$ outcome labels.  No new numerical-relativity
evolutions are introduced here; the existing catalogue is used as a fixed
data set for training, validation, and held-out testing.

We construct branch-conditioned reduced-basis surrogates for the real
$(\ell,m)=(2,0)$ and $(4,0)$ waveform modes.  For each mode, the continuous
parameters $(|\phi_c|,\lambda)$ are mapped to a branch-specific reduced
representation and then reconstructed into the corresponding waveform.
Neural reduced-order and surrogate waveform models are well established in
gravitational-wave modelling~\cite{Cuoco:2024cdk,Chua:2018woh}; here the
surrogate is used primarily as a generative tool for testing whether the
merger branch and source parameters can be identified from waveform
morphology.

The analysis is performed at two levels.  First, we assess the forward
waveform reconstruction and the separation of the merger branches in waveform
space on the frozen benchmark.  Second, we perform parameter-blind inference
in which the true values of $|\phi_c|$, $\lambda$, and the merger branch are
withheld from the inference procedure.  Candidate waveforms are compared
directly with each held-out numerical-relativity signal, allowing the branch
and intrinsic parameters to be inferred from the waveform itself.  The
additional $(4,0)$ mode provides a separate test of whether the branch
information is confined to the dominant mode.

On the frozen held-out test set, the three merger branches remain
separable using either the $(2,0)$ or the $(4,0)$ mode, while the
parameter-blind analysis also recovers the merger branch without supplying
the intrinsic source parameters.  We further quantify the recovery of
$|\phi_c|$ and $\lambda$, and examine the branch dependence of the resulting
parameter errors.

The scope remains restricted to the existing equal-mass, head-on,
axisymmetric catalogue.  The present analysis includes the $(2,0)$ and
$(4,0)$ modes, but not generic orbital configurations, unequal masses,
detector response, or measurement noise.  The results should therefore be
interpreted as a waveform-level inference test within this catalogue rather
than as a detector-ready classifier for generic boson-star binaries.

The paper is organized as follows.  Section~\ref{sec:data} summarizes the
numerical-relativity catalogue and multimode data set.
Section~\ref{sec:method} introduces the branch-conditioned reduced-basis
surrogate, and Sec.~\ref{sec:metrics} describes the waveform comparison and
parameter-blind inference procedure.  The results are presented in
Sec.~\ref{sec:results}, followed by the discussion in
Sec.~\ref{sec:discussion} and conclusions in
Sec.~\ref{sec:conclusion}.  Throughout this work we use geometric units with
$G=c=1$.

\section{Numerical-relativity waveform catalogue and multimode data set}
\label{sec:data}

All gravitational-wave data used in this paper are taken directly from the
numerical-relativity catalogue of Ref.~\cite{Ge:2025btw}.  The boson-star
solutions, binary initial data, merger-outcome labels, {\sc GRChombo}-based
Cartoon evolutions~\cite{Andrade:2021rbd,Ge:2024fum,Cook:2016soy,
Cook:2016qnt}, gravitational-wave extraction, and convergence tests are those
described in Refs.~\cite{Ge:2024itl,Ge:2025btw}.  No new
numerical-relativity evolutions are introduced here; the existing waveforms
and branch labels are used for surrogate construction and held-out testing.

\subsection{Massive boson-star catalogue}
\label{subsec:bs_catalogue}

The source catalogue consists of head-on equal-mass mergers of quartically
self-interacting massive boson stars.  The matter sector is a complex scalar
field minimally coupled to gravity, with stationary isolated configurations
written as
\begin{equation}
\phi(t,r)=|\phi(r)|e^{i\omega t}.
\end{equation}
The scalar-field potential is
\begin{equation}
V(|\phi|^2)
=
m^2|\phi|^2
+
\frac{\lambda}{2}|\phi|^4 ,
\label{eq:quartic_potential}
\end{equation}
where $m$ is the scalar-field mass and $\lambda$ is the quartic
self-interaction strength.  For fixed $\lambda$, the equilibrium sequence is
labelled by the central scalar amplitude $|\phi_c|$.

The analysis uses the four catalogue values
$\lambda\in\{10,50,100,300\}$ and includes all three merger-outcome branches
in a common benchmark.

\subsection{Merger-outcome branches}
\label{subsec:branches}

The head-on evolutions in Ref.~\cite{Ge:2025btw} exhibit three qualitatively
different merger outcomes.  These outcomes define the branch structure used
throughout the present paper.

The branch $\mathrm{BS}_{\rm post}$ contains mergers that remain
horizonless over the simulated interval and form a boson-star remnant.
In $\mathrm{BH}_{\rm post}$, the scalar configurations first come into
contact and subsequently collapse to a black hole.  In
$\mathrm{BH}_{\rm pre}$, each star collapses before contact, so that the
late-time evolution is effectively a head-on black-hole binary coalescence.

We denote the branch set by
\begin{equation}
\mathcal{B}
=
\left\{
\mathrm{BS}_{\rm post},
\mathrm{BH}_{\rm post},
\mathrm{BH}_{\rm pre}
\right\}.
\label{eq:branch_set}
\end{equation}
The label $s\in\mathcal{B}$ selects the corresponding branch-specific head of
the surrogate described in Sec.~\ref{sec:method}.

\subsection{Waveform modes and data split}
\label{subsec:waveform_data}

Axisymmetry restricts the radiation to $m=0$ modes.  We use the real
components of the $(\ell,m)=(2,0)$ and $(4,0)$ modes, denoted by
$h_{20}^{\rm NR}(t)$ and $h_{40}^{\rm NR}(t)$, respectively.  These symbols
refer to the processed extracted waveform channels of the numerical-relativity
catalogue rather than to an additional strain reconstruction.  Separate
surrogate instances with the same reduced-basis architecture are trained for
the two modes, and the processed waveforms are represented on a common time
grid with 1024 samples.

The 415 catalogue samples are divided into 331 training, 24 validation,
and 60 test samples.  The test set is balanced across the four values of
$\lambda$ and the three merger branches, with five samples for each
$(\lambda,s)$ combination.  The split, reduced-basis dimensions, model
architecture, and inference prescription are fixed before evaluation on the
test set.

\section{Branch-conditioned reduced-basis waveform surrogate}
\label{sec:method}

We construct a forward surrogate for each of the real $(2,0)$ and $(4,0)$
waveform modes.  Rather than predicting the full time series directly, the
surrogate separates the overall waveform scale from its shape and represents
the latter in a branch-dependent reduced basis.  This gives a compact
description of the three waveform families while retaining their distinct
merger dynamics.

\subsection{Reduced-basis waveform representation}
\label{subsec:reduced_basis}

For either mode, let $h_{\ell0}^{\rm NR}(t_j)$ denote the processed
numerical-relativity waveform, with $\ell=2$ or $4$,
$j=1,\ldots,N_t$, and $N_t=1024$.  We define its root-mean-square
amplitude by
\begin{equation}
R_{\ell0}
=
\left[
\frac{1}{N_t}
\sum_{j=1}^{N_t}
\left(h_{\ell0}^{\rm NR}(t_j)\right)^2
\right]^{1/2},
\label{eq:rms_amplitude}
\end{equation}
and separate the waveform shape as
\begin{equation}
g_{\ell0}(t_j)
=
\frac{h_{\ell0}^{\rm NR}(t_j)}
{R_{\ell0}} .
\label{eq:normalised_shape}
\end{equation}
A fixed numerical floor is used in the implementation to avoid division by,
or the logarithm of, a vanishing amplitude.

The waveform shapes are reduced separately for each merger branch.  For
$s\in\mathcal{B}$, principal-component analysis (PCA) of the corresponding
training waveforms gives
\begin{equation}
g_{\ell0}(t)
\simeq
\mu_{\ell s}(t)
+
\sum_{k=1}^{K}
c_{\ell s,k}\,
e_{\ell s,k}(t),
\label{eq:pca_representation}
\end{equation}
where $\mu_{\ell s}$ is the branch mean, $e_{\ell s,k}$ are the basis
functions, and $c_{\ell s,k}$ are the reduced coefficients.  We use
$K=16$ for both modes, fixed before evaluation on the test set.

Each waveform is therefore represented by
\begin{equation}
\mathbf{q}_{\ell s}
=
\left(
\log R_{\ell0},
c_{\ell s,1},
\ldots,
c_{\ell s,K}
\right).
\label{eq:latent_representation}
\end{equation}
After predicting these quantities, the physical waveform is reconstructed as
\begin{equation}
h_{\ell0}^{\rm sur}(t)
=
\exp\!\left(\widehat{\log R}_{\ell0}\right)
\left[
\mu_{\ell s}(t)
+
\sum_{k=1}^{K}
\widehat{c}_{\ell s,k}\,
e_{\ell s,k}(t)
\right].
\label{eq:waveform_reconstruction}
\end{equation}

\subsection{Branch-conditioned parameter map}
\label{subsec:branch_map}

The surrogate implements the forward map
\begin{equation}
\left(|\phi_c|,\lambda,s\right)
\longmapsto
\mathbf{q}_{\ell s}
\longmapsto
h_{\ell0}^{\rm sur}(t).
\label{eq:surrogate_map}
\end{equation}
The continuous inputs are represented by $(|\phi_c|,\log\lambda)$ and
standardised using training-set statistics.  They are passed through a shared
encoder, while the branch label $s$ selects one of three branch-specific
output heads.  Thus the branch label selects the appropriate waveform
manifold rather than being treated as a continuous input parameter.

The network predicts the standardised form of
Eq.~\eqref{eq:latent_representation}.  The shared encoder has three
fully connected layers of width 128, and each branch head contains one
hidden layer of width 128 followed by the $K+1=17$ output quantities.
Separate instances of the same architecture are trained for the $(2,0)$ and
$(4,0)$ modes.

\begin{figure}[t]
    \centering
    \includegraphics[width=0.75\columnwidth]
    {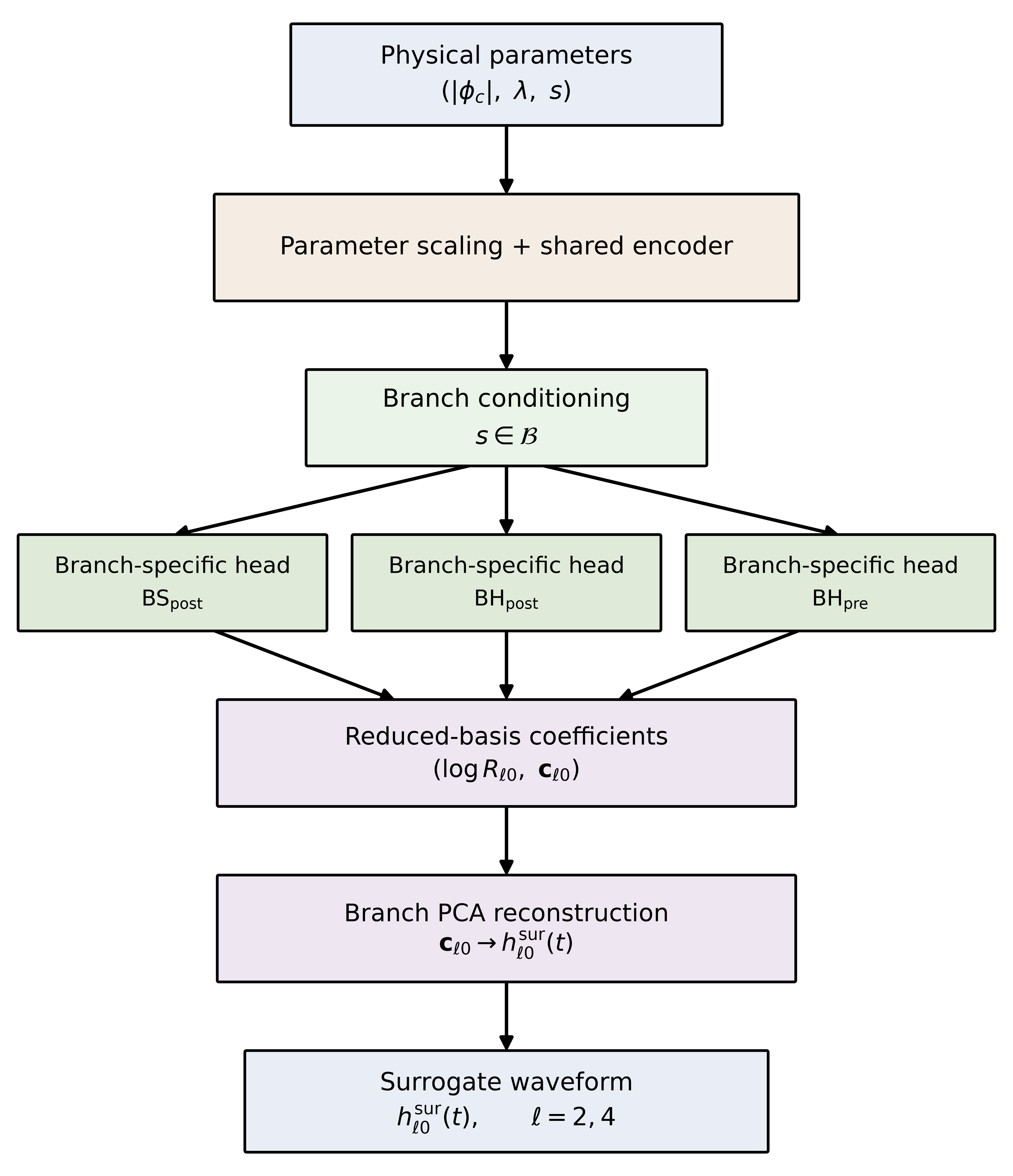}
    \caption{
    Forward construction of the branch-conditioned reduced-basis surrogate.
    The continuous source parameters are passed through a shared encoder,
    while the merger branch selects the corresponding output head.  The head
    predicts the logarithmic RMS amplitude and the branch-specific PCA
    coefficients.  The PCA shape reconstruction is then rescaled by the
    predicted RMS amplitude to obtain the physical waveform.  Separate
    instances of the same architecture are used for the real $(2,0)$ and
    $(4,0)$ modes.
    }
    \label{fig:surrogate_architecture}
\end{figure}

Figure~\ref{fig:surrogate_architecture} summarises this construction.  The
essential feature is the branch-conditioned reduced representation: the
network interpolates the dependence of the waveform scale and shape on the
source parameters, while the branch-specific basis retains the distinct
dynamical waveform families.

\subsection{Training and model selection}
\label{subsec:training}

The parameter scaling, branch-specific PCA bases, and latent-variable
standardisation are determined from the training set only.  The latent
variables are standardised separately within each branch before training.

For a mini-batch of $N_{\rm b}$ samples, let $a_i$ denote the standardised
logarithmic RMS amplitude and $c_{i,k}$ the standardised PCA coefficients.
The corresponding network predictions are denoted by
$\widehat{a}_i$ and $\widehat{c}_{i,k}$.  The training objective is
\begin{equation}
\mathcal{L}_{\rm latent}
=
\frac{1}{N_{\rm b}}
\sum_{i=1}^{N_{\rm b}}
\left(\widehat{a}_i-a_i\right)^2
+
\frac{1}{N_{\rm b}K}
\sum_{i=1}^{N_{\rm b}}
\sum_{k=1}^{K}
\left(\widehat{c}_{i,k}-c_{i,k}\right)^2 .
\label{eq:latent_loss}
\end{equation}
The amplitude and shape contributions therefore enter with equal weight.

Training uses branch-balanced sampling so that the more densely populated
branches do not dominate the optimisation.  The networks are trained with
AdamW for at most 1500 epochs, with batch size 64, initial learning rate
$10^{-3}$, weight decay $10^{-5}$, cosine learning-rate decay, and early
stopping with patience 180.  Checkpoint selection is based on the validation
loss, and no test-set information enters the training or model selection.

The canonical waveform alignment used to construct the reduced bases is
determined from the training data only.  In parameter-blind inference, the
remaining relative timing freedom of the held-out target is treated as a
nuisance parameter and profiled over, as described in
Sec.~\ref{sec:metrics}.

\section{Waveform comparison and parameter-blind inference}
\label{sec:metrics}

The forward surrogate of Sec.~\ref{sec:method} associates each candidate
source point with a waveform.  We use this map for inference by comparing
surrogate waveforms with a held-out numerical-relativity signal.  The true
source parameters and merger branch of the target are not used to construct
the candidate set.

\subsection{Waveform score}
\label{subsec:waveform_metrics}

For a candidate source hypothesis
\begin{equation}
\vartheta=(|\phi_c|,\lambda,s),
\qquad
s\in\mathcal{B},
\end{equation}
let $h_{\ell0}^{\rm sur}(t;\vartheta)$ denote the corresponding surrogate
waveform and $h_{\ell0}^{\rm NR}(t)$ the held-out target, with
$\ell=2$ or $4$.

We use the relative $L_2$ error
\begin{equation}
\epsilon_{\ell}(\vartheta)
=
\frac{
\left\|
h_{\ell0}^{\rm sur}(\vartheta)
-
h_{\ell0}^{\rm NR}
\right\|_2
}{
\left\|
h_{\ell0}^{\rm NR}
\right\|_2
+
\epsilon_{\rm met}
},
\label{eq:relative_l2}
\end{equation}
together with the absolute normalised overlap
\begin{equation}
\mathcal{O}_{\ell}(\vartheta)
=
\frac{
\left|
\left\langle
h_{\ell0}^{\rm sur}(\vartheta),
h_{\ell0}^{\rm NR}
\right\rangle
\right|
}{
\left\|
h_{\ell0}^{\rm sur}(\vartheta)
\right\|_2
\left\|
h_{\ell0}^{\rm NR}
\right\|_2
+
\epsilon_{\rm met}
}.
\label{eq:waveform_overlap}
\end{equation}
The first quantity retains sensitivity to the waveform amplitude, whereas the
second mainly measures its shape.  We combine them through
\begin{equation}
Q_{\ell}(\vartheta)
=
\frac{1}{2}
\left[
\mathcal{O}_{\ell}(\vartheta)
+
\frac{1}{1+\epsilon_{\ell}(\vartheta)}
\right].
\label{eq:hybrid_score}
\end{equation}
Branch-given reconstruction uses these quantities at the true catalogue
point in the canonical waveform frame.

For parameter-blind inference, the residual arrival-time offset is treated as
a nuisance parameter.  For a relative shift $\Delta t$, the score is
evaluated on the overlapping part of the two waveforms, and we define
\begin{equation}
S_{\ell}(\vartheta)
=
\max_{|\Delta t|\leq\Delta t_{\max}}
Q_{\ell}(\vartheta;\Delta t).
\label{eq:profiled_score}
\end{equation}
We use
\begin{equation}
\Delta t_{\max}=60
\end{equation}
in the time coordinate of the processed waveforms and require at least
$75\%$ of the waveform window to remain in the comparison.

\subsection{Parameter-blind search}
\label{subsec:blind_search}

The candidate bank is determined entirely from the training catalogue.
For every training-supported pair $(\lambda,s)$, the interval in
$|\phi_c|$ covered by the corresponding training samples is populated by
$N_{\phi}=250$ uniformly spaced candidates.  The bank therefore has the form
\begin{equation}
\Theta_{\rm bank}
=
\bigcup_{(\lambda,s)}
\left\{
\left(|\phi_c|_n,\lambda,s\right)
\right\}_{n=1}^{N_{\phi}},
\label{eq:candidate_bank}
\end{equation}
with
\begin{equation}
\lambda\in\{10,50,100,300\}.
\end{equation}
Neither the true parameters nor the branch label of a validation or test
sample enter this construction.

For a given mode and network seed $r$, the best candidate is
\begin{equation}
\widehat{\vartheta}_{\ell}^{(r)}
=
\arg\max_{\vartheta\in\Theta_{\rm bank}}
S_{\ell}^{(r)}(\vartheta).
\label{eq:single_mode_inference}
\end{equation}
This returns a merger branch, a value of $|\phi_c|$, and one of the four
catalogue values of $\lambda$.  The present recovery of $\lambda$ is therefore
discrete rather than a continuous estimate of the self-interaction strength.

The fixed-parameter comparison of the original proof of concept corresponds
to restricting the search to a specified $(|\phi_c|,\lambda)$.  The revised
analysis removes this restriction and searches the full training-supported
candidate bank.

For branch diagnostics, the score is first maximised over the source
parameters within each candidate branch.  If $S^{(1)}$ and $S^{(2)}$ denote
the largest and second-largest of these branch-wise maxima, we define
\begin{equation}
\Delta S
=
S^{(1)}-S^{(2)}.
\label{eq:branch_margin}
\end{equation}

\subsection{Multimode comparison}
\label{subsec:multimode_inference}

We consider the $(2,0)$ and $(4,0)$ modes separately and in combination.
For the symmetric multimode benchmark, both modes must correspond to the same
candidate source point and the same residual time shift.  For each seed,
\begin{equation}
Q_{20+40}^{(r)}(\vartheta;\Delta t)
=
\frac{1}{2}
\left[
Q_{20}^{(r)}(\vartheta;\Delta t)
+
Q_{40}^{(r)}(\vartheta;\Delta t)
\right],
\label{eq:joint_score}
\end{equation}
and
\begin{equation}
S_{20+40}^{(r)}(\vartheta)
=
\max_{|\Delta t|\leq\Delta t_{\max}}
Q_{20+40}^{(r)}(\vartheta;\Delta t).
\label{eq:joint_profiled_score}
\end{equation}
The equal mode weights are fixed before evaluation on the test set.

Two simple controls are also used.  In the shape-only test,
Eq.~\eqref{eq:hybrid_score} is replaced by the normalised overlap alone.
In the nearest-template test, the surrogate candidate waveforms are replaced
by the corresponding training waveforms.  These controls test whether the
branch separation is already present in waveform morphology rather than being
created by a particular surrogate representation.

\subsection{Frozen final inference rule}
\label{subsec:final_inference}

The final estimator is fixed using the validation set and is deliberately
asymmetric.  The merger branch and $\lambda$ are obtained from the standalone
$(4,0)$ search.  With five independently trained seeds,
\begin{equation}
\widehat{s}
=
{\rm Maj}_{r}
\left[
\widehat{s}_{40}^{(r)}
\right],
\qquad
\widehat{\lambda}
=
{\rm Maj}_{r}
\left[
\widehat{\lambda}_{40}^{(r)}
\right],
\label{eq:final_branch_lambda}
\end{equation}
where ${\rm Maj}$ denotes the majority value.

For $|\phi_c|$, the residual time shift is first profiled independently for
each mode.  For seed $r$ we define
\begin{equation}
F_{w}^{(r)}(\vartheta)
=
w\,S_{20}^{(r)}(\vartheta)
+
(1-w)\,S_{40}^{(r)}(\vartheta).
\label{eq:late_fusion_score}
\end{equation}

The standalone $(4,0)$ gate is also used to group the validation samples when
selecting the fusion weight.  Thus the weight selection depends on the
inferred branch, not on the true branch label of the target.  Within each
gated validation subset, $w$ is chosen to minimise the mean relative error in
$|\phi_c|$.  The selected $(2,0)$ weights are
\begin{equation}
w_{\mathrm{BS}_{\rm post}}=0.95,
\qquad
w_{\mathrm{BH}_{\rm post}}=0.75,
\qquad
w_{\mathrm{BH}_{\rm pre}}=0.10.
\label{eq:phic_fusion_weights}
\end{equation}

For each seed, the weight selected by the final branch gate is applied to the
full candidate bank,
\begin{equation}
\widehat{\vartheta}_{\phi}^{(r)}
=
\arg\max_{\vartheta\in\Theta_{\rm bank}}
F_{w_{\widehat{s}}}^{(r)}(\vartheta),
\label{eq:phic_candidate}
\end{equation}
and $\widehat{|\phi_c|}^{(r)}$ is the $|\phi_c|$ component of
$\widehat{\vartheta}_{\phi}^{(r)}$.  The final estimate is the five-seed
average,
\begin{equation}
\widehat{|\phi_c|}
=
\frac{1}{5}
\sum_{r=1}^{5}
\widehat{|\phi_c|}^{(r)} .
\label{eq:final_phic}
\end{equation}
No test sample is used to select the fusion weights or any other part of the
inference prescription.

This late-fusion estimator differs from
Eq.~\eqref{eq:joint_profiled_score}.  The symmetric multimode benchmark uses
one common nuisance shift for both modes, whereas the final $|\phi_c|$
estimator profiles the two modes separately before combining their scores.
Both procedures are fixed before the held-out test set is evaluated.

\section{Results}
\label{sec:results}

We evaluate the surrogate and the parameter-blind inference procedure on the
frozen set of 60 held-out numerical-relativity samples.  Each sample provides
both the real $(2,0)$ and $(4,0)$ waveform modes.  The surrogate benchmark is
repeated over five independently trained network seeds, while the final
inverse estimator combines these seeds according to the prescription of
Sec.~\ref{subsec:final_inference}.  We first examine waveform reconstruction
and branch diagnosis, and then turn to the recovery of the source parameters.

\subsection{Multimode waveform reconstruction and branch diagnosis}
\label{subsec:multimode_results}

Table~\ref{tab:multimode_branch_results} summarises the multimode benchmark.
The waveform overlap and relative-$L_2$ error refer to branch-given
reconstruction at the true catalogue point.  The branch accuracies are instead
obtained from the parameter-blind search of
Sec.~\ref{subsec:blind_search}.  Uncertainties are sample standard deviations
over the five network seeds.

\begin{table*}[t]
\centering
\begin{tabular}{lccccc}
\hline\hline
Waveform input
& $\langle\mathcal{O}\rangle$
& $\langle\epsilon_{L_2}\rangle$
& Hybrid branch accuracy
& Shape-only accuracy
& $\langle\Delta S\rangle$
\\
\hline
$(2,0)$
& $0.99749\pm0.00027$
& $0.0697\pm0.0066$
& $60/60$
& $60/60$
& $0.2089\pm0.0008$
\\
$(4,0)$
& $0.91447\pm0.00195$
& $0.2937\pm0.0095$
& $60/60$
& $60/60$
& $0.2523\pm0.0036$
\\
$(2,0)+(4,0)$
& ---
& ---
& $60/60$
& $60/60$
& $0.2497\pm0.0029$
\\
\hline\hline
\end{tabular}
\caption{
Performance on the frozen held-out test set.  The first two columns give the
mean branch-given reconstruction overlap and relative-$L_2$ error for the
single-mode surrogates.  The remaining columns give the parameter-blind branch
accuracy using the hybrid and shape-only scores, together with the mean
hybrid-score separation between the preferred branch and its nearest
competitor.  Reconstruction statistics and margins are means over five
independently trained network seeds, with sample standard deviations shown.
Each $60/60$ branch accuracy is obtained for every seed.  The combined
$(2,0)+(4,0)$ entry is a multimode score rather than a single reconstructed
waveform, so no joint reconstruction metric is quoted.
}
\label{tab:multimode_branch_results}
\end{table*}

The dominant $(2,0)$ mode is reconstructed accurately, with a mean overlap of
$0.99749$ and a mean relative-$L_2$ error of $0.0697$.  The $(4,0)$ mode is
more difficult to reconstruct point by point, with a mean overlap of $0.91447$
and a mean relative-$L_2$ error of $0.2937$.  This lower reconstruction
fidelity does not prevent branch diagnosis: the parameter-blind hybrid search
recovers all 60 branch labels from the $(4,0)$ mode alone for every network
seed.

The same $60/60$ result is obtained from the $(2,0)$ mode and from the
common-shift multimode score.  It is also unchanged when the hybrid score is
replaced by the normalised overlap alone.  Since the $(2,0)$ mode already
gives perfect branch recovery on this test set, the $(4,0)$ result should not
be interpreted as an improvement in classification accuracy.  Rather, it
shows that the branch information is not confined to the dominant mode.

A nearest-template control gives the same conclusion without neural
interpolation.  Direct comparison with the training waveforms gives $60/60$
correct branch assignments for the $(2,0)$ mode, the $(4,0)$ mode, and their
combination, for both the hybrid and shape-only scores.  The corresponding
mean hybrid margins are $0.193$, $0.265$, and $0.247$, respectively.  The
branch separation is therefore already present in waveform space rather than
being generated by the surrogate architecture.

Representative branch-given reconstructions of both modes are shown in
Appendix~\ref{app:reconstructions}. They provide a waveform-level check of the reduced-basis
reconstruction, while the main analysis below concerns branch and
source-parameter inference.

\subsection{Parameter-blind recovery of the source parameters}
\label{subsec:parameter_recovery}

We next apply the frozen final inference rule of
Sec.~\ref{subsec:final_inference}.  The numerical-relativity waveform is
supplied without its true $|\phi_c|$, $\lambda$, or merger branch.  The
candidate bank and all validation-selected choices are fixed before the test
set is evaluated.

The standalone $(4,0)$ branch gate recovers all 60 held-out branch labels.
Perfect branch recovery is therefore retained without supplying the intrinsic
source parameters.

\begin{figure}[t]
\centering
\includegraphics[width=0.75\columnwidth]{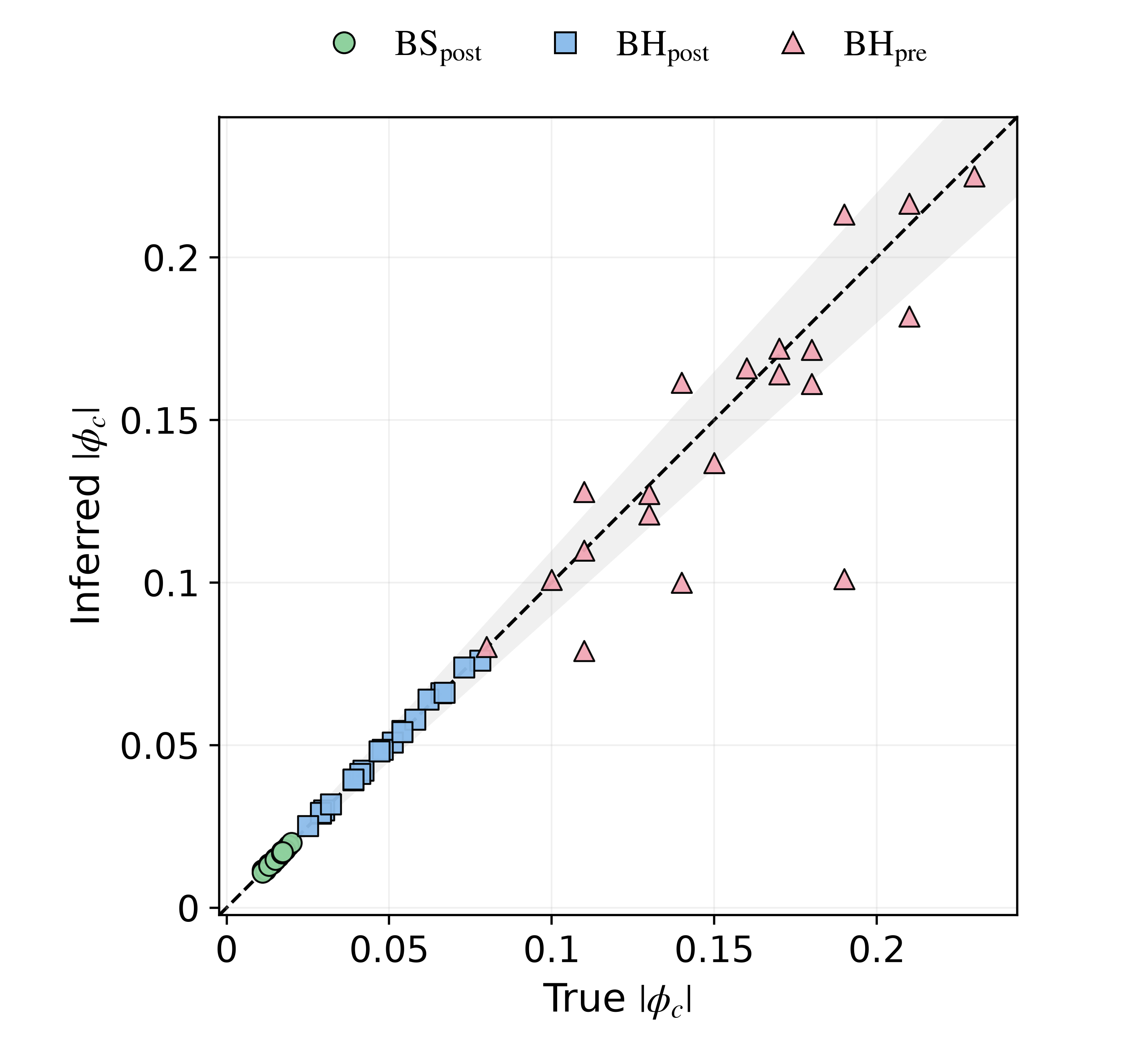}
\caption{
Parameter-blind recovery of the central scalar amplitude on the frozen
held-out test set.  The horizontal axis gives the true numerical-relativity
value and the vertical axis the inferred value.  Markers distinguish the true
merger branches.  The dashed line denotes exact recovery, and the shaded
region corresponds to a relative error of $\pm10\%$.  The true source
parameters are used only to assess the recovery and are not supplied during
inference.
}
\label{fig:phic_recovery}
\end{figure}

Figure~\ref{fig:phic_recovery} shows the recovery of the central scalar
amplitude.  Across the full test set, the mean absolute error is
$0.005735$, while the median absolute error is
$3.74\times10^{-4}$.  The mean and median relative errors are $4.30\%$ and
$1.31\%$, respectively.  The inferred value lies within $5\%$, $10\%$, and
$20\%$ of the true value for $83.3\%$, $86.7\%$, and $95.0\%$ of the test
samples.

\begin{figure}[t]
\centering
\includegraphics[width=0.55\columnwidth]{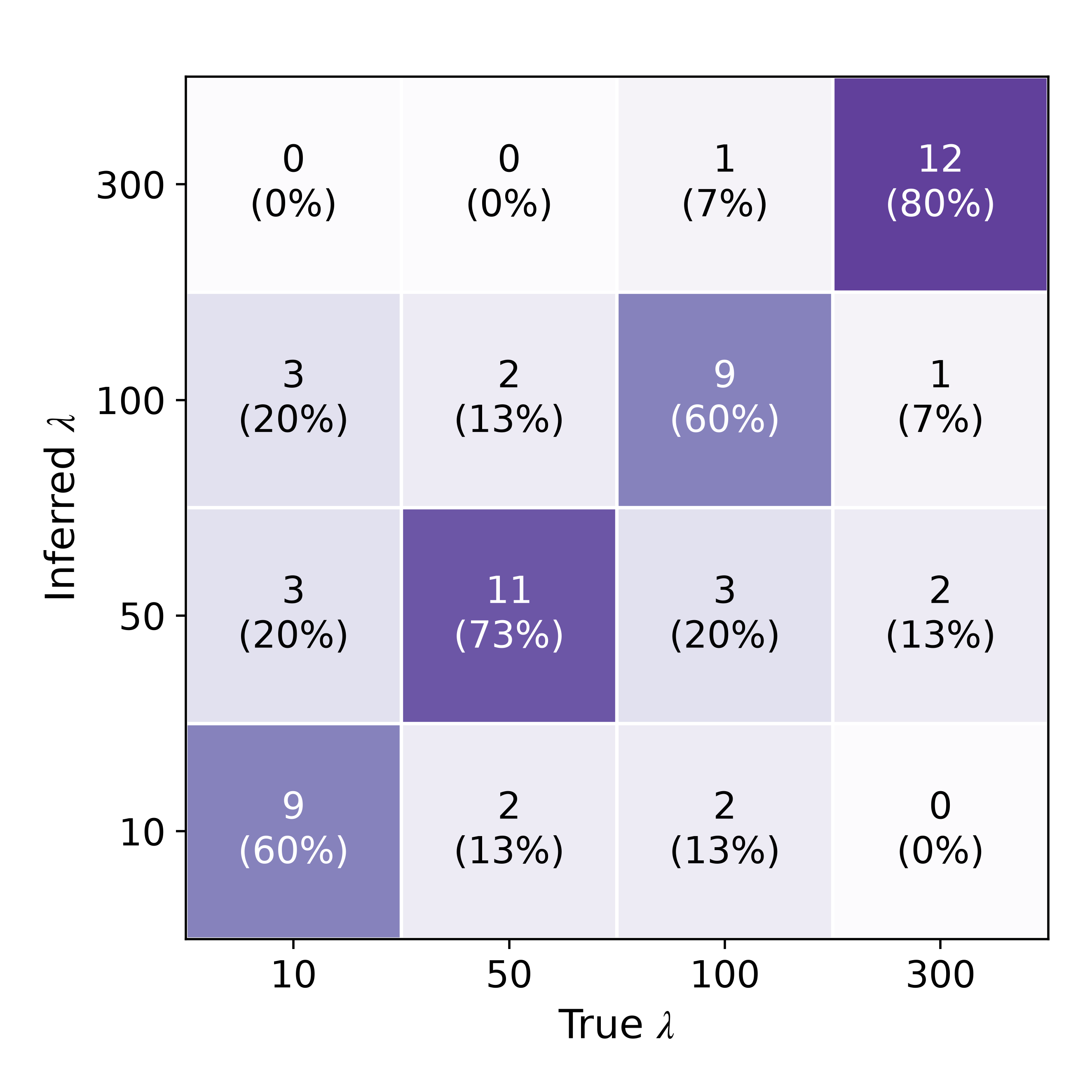}
\caption{
Parameter-blind recovery of the self-interaction strength on the frozen
held-out test set.  Columns correspond to the true catalogue value of
$\lambda$, while rows give the inferred value.  Each cell shows the number
of test samples and the percentage within the corresponding true-$\lambda$
column.  The inferred value is selected from
$\lambda\in\{10,50,100,300\}$ using the standalone $(4,0)$ waveform score.
}
\label{fig:lambda_confusion}
\end{figure}

Figure~\ref{fig:lambda_confusion} shows the corresponding recovery of the
self-interaction strength.  In the present catalogue, $\lambda$ is selected
from the four discrete values $\{10,50,100,300\}$.  The exact value is
recovered for $41/60=68.3\%$ of the test samples, while
$46/60=76.7\%$ are recovered within a factor of two.  The mean absolute
error in $\log\lambda$ is $0.480$.  These results refer to the standalone
$(4,0)$ estimator used in the final inference rule and should not be
interpreted as a continuous measurement of $\lambda$.

\subsection{Branch dependence of the parameter recovery}
\label{subsec:branchwise_recovery}

The source-parameter errors show a substantial dependence on the merger
branch.  We define the relative central-amplitude error as
\begin{equation}
\delta_{\phi}
=
\frac{
\left|
\widehat{|\phi_c|}
-
|\phi_c|_{\rm true}
\right|
}{
|\phi_c|_{\rm true}
}.
\label{eq:phic_relative_error}
\end{equation}
The branch-resolved results are given in
Table~\ref{tab:branchwise_parameter_recovery}.

\begin{table*}[t]
\centering
\begin{tabular}{lcccccccc}
\hline\hline
Branch
& $N$
& Branch acc.
& $\langle\delta_{\phi}\rangle$
& $\delta_{\phi}<5\%$
& $\delta_{\phi}<10\%$
& $\delta_{\phi}<20\%$
& $\lambda$ exact
& $\lambda$ within factor 2
\\
\hline
$\mathrm{BS}_{\rm post}$
& $20$
& $100\%$
& $1.54\%$
& $100\%$
& $100\%$
& $100\%$
& $70\%$
& $80\%$
\\
$\mathrm{BH}_{\rm post}$
& $20$
& $100\%$
& $0.91\%$
& $100\%$
& $100\%$
& $100\%$
& $100\%$
& $100\%$
\\
$\mathrm{BH}_{\rm pre}$
& $20$
& $100\%$
& $10.44\%$
& $50\%$
& $60\%$
& $85\%$
& $35\%$
& $50\%$
\\
\hline
All
& $60$
& $100\%$
& $4.30\%$
& $83.3\%$
& $86.7\%$
& $95.0\%$
& $68.3\%$
& $76.7\%$
\\
\hline\hline
\end{tabular}
\caption{
Branch-resolved performance of the frozen parameter-blind inference.
The central scalar amplitude is assessed by its mean relative error and by
the fraction of samples within three relative-error thresholds.  The
self-interaction strength is assessed by exact recovery and recovery within
a factor of two.  Each branch contains 20 held-out samples.
}
\label{tab:branchwise_parameter_recovery}
\end{table*}

The distinction between branch identification and source-parameter recovery
is clear.  The branch is recovered for every held-out sample, but the
remaining parameter error is concentrated mainly in
$\mathrm{BH}_{\rm pre}$.  For $\mathrm{BH}_{\rm post}$, the mean relative
error in $|\phi_c|$ is $0.91\%$, and $\lambda$ is recovered exactly for all
20 samples.  For $\mathrm{BS}_{\rm post}$, the corresponding
$|\phi_c|$ error is $1.54\%$, while the exact-$\lambda$ accuracy is $70\%$.

The $\mathrm{BH}_{\rm pre}$ branch is substantially more difficult.  Its
mean relative $|\phi_c|$ error is $10.44\%$, only $60\%$ of the samples lie
within $10\%$ of the true value, and the exact-$\lambda$ recovery falls to
$35\%$.  Thus, on the present test set, the main limitation of the
parameter-blind inference is source-parameter recovery within the
$\mathrm{BH}_{\rm pre}$ waveform family rather than identification of the
merger branch.  We return to this point in Sec.~\ref{sec:discussion}.

\section{Discussion}
\label{sec:discussion}

\subsection{Waveform morphology and merger-route information}
\label{subsec:discussion_morphology}

The main result is that the dynamical merger route is encoded in the
gravitational waveform strongly enough to support parameter-blind branch
identification within the present numerical-relativity catalogue.  The three
branches were originally defined from the nonlinear evolution itself:
whether the system remains horizonless, collapses after contact, or collapses
before contact.  The present analysis shows that the same distinction is also
visible in waveform space.

This point is clearest for the two black-hole-forming branches.
Both $\mathrm{BH}_{\rm post}$ and $\mathrm{BH}_{\rm pre}$ end in black-hole
formation, but they reach that state through different dynamics.  In the
former, the scalar configurations interact before collapse; in the latter,
the individual stars collapse before contact.  Their successful separation
from the processed numerical-relativity waveform therefore shows that the
signal retains information about the route to the final state, rather than
only the nature of the final remnant.

The higher-mode result strengthens this interpretation.  The real $(4,0)$
mode is reconstructed less accurately than the dominant $(2,0)$ mode, yet it
still recovers all three branches across the frozen test set.  This should not
be interpreted as an improvement in classification accuracy, since the
$(2,0)$ mode already gives perfect branch recovery.  Instead, it shows that
the branch information is not confined to the dominant mode.  The $(4,0)$
waveform provides a separate multipolar channel in which the same merger-route
information remains visible despite the greater reconstruction difficulty.

The nearest-template control leads to the same branch assignments without
neural interpolation.  The separation is therefore already present in the
numerical-relativity waveform manifold.  The role of the surrogate is to
provide a smooth branch-conditioned forward model over the training-supported
parameter domain and to make that structure available for inverse waveform
comparison.  The branch information itself is not created by the network.

\subsection{Parameter-blind inference and branch-dependent ambiguity}
\label{subsec:discussion_inference}

The revised inference problem is more demanding than the fixed-parameter
comparison of the original proof of concept.  The true $|\phi_c|$, $\lambda$,
and merger branch are not supplied during candidate scoring.  The processed
numerical-relativity waveform is instead compared with a bank of source
hypotheses constructed entirely from the training catalogue.  The successful
branch recovery therefore shows that the merger route can be identified
without first supplying the intrinsic source parameters.

Branch identification and source-parameter recovery are nevertheless not
equally difficult.  The branch label is recovered for every held-out sample,
whereas the errors in $|\phi_c|$ and $\lambda$ vary substantially between
branches.  Most of the remaining parameter error is concentrated in
$\mathrm{BH}_{\rm pre}$, while $\mathrm{BH}_{\rm post}$ and
$\mathrm{BS}_{\rm post}$ are considerably better constrained.

The poorer recovery within $\mathrm{BH}_{\rm pre}$ points to a stronger
ambiguity in the mapping from source parameters to waveform morphology in
this part of the catalogue.  Once the stars collapse before contact, the
subsequent signal can become dominated by black-hole-like dynamics, so
different progenitor configurations may produce similar late-time
waveforms.  However, the present analysis does not establish that this is the
sole origin of the observed errors.  Local catalogue coverage and surrogate
interpolation accuracy may also contribute.  Distinguishing these effects
would require a denser numerical-relativity catalogue and a direct comparison
between waveform variation and quantities such as collapse time, compactness,
and remnant properties.

The asymmetric final estimator reflects the different parameter-recovery
performance of the two waveform modes across the present catalogue.  The
$(4,0)$ mode provides the frozen branch gate and the estimate of $\lambda$,
whereas the final $|\phi_c|$ estimate combines the independently profiled
$(2,0)$ and $(4,0)$ scores using validation-selected weights.  This is a
practical inference rule selected for the present data set, rather than
evidence that one mode is generically more informative than the other.  Their
relative usefulness may change for more general binaries or once detector
effects are included.

\subsection{Scope and limitations}
\label{subsec:discussion_limitations}

The present results apply to interpolation within the existing
training-supported catalogue.  The binaries are equal-mass, head-on, and
axisymmetric, and the analysis uses only the real $(2,0)$ and $(4,0)$ modes.
The calculation therefore does not address unequal masses, nonzero impact
parameter, orbital angular momentum, or generic $m\neq0$ radiation.  These
extensions would enlarge both the physical parameter space and the waveform
manifold, and it is not yet known whether the same branch separation would
remain equally clean.

The parameter-blind test is also performed directly on processed
numerical-relativity waveforms.  Detector response and detector noise are not
included, nor is a likelihood-based posterior analysis.  The $100\%$ branch
accuracy reported here should therefore be interpreted only as a result for
the frozen noiseless numerical-relativity test set.  It is not an estimate of
the classification accuracy achievable in gravitational-wave observations.

A further limitation concerns $\lambda$.  The present estimator selects
among the four catalogue values
$\{10,50,100,300\}$ and does not infer a continuous self-interaction
strength.  Likewise, the candidate values of $|\phi_c|$ are restricted to
the parameter intervals represented in the training catalogue.  The current
analysis therefore tests held-out recovery within known support rather than
extrapolation to arbitrary massive-boson-star configurations.

These limitations also indicate the most useful next steps.  A denser and
more general numerical-relativity catalogue would allow the branch structure
to be tested away from head-on symmetry and would help separate genuine
waveform degeneracies from limitations of local catalogue coverage.  Adding
detector response and noise would then determine whether the waveform-space
separation survives under observational conditions.  These extensions are
more important for establishing the physical reach of the method than further
optimisation of the present network architecture.

Within its present scope, the conclusion is nevertheless clear.  The three
dynamical merger routes identified from the nonlinear evolutions are also
distinguishable in the emitted gravitational radiation.  This information is
present in both the real $(2,0)$ and $(4,0)$ modes and supports
parameter-blind branch identification on held-out numerical-relativity
waveforms.  The remaining difficulty lies mainly in source-parameter recovery
within part of the waveform catalogue, rather than in identifying the merger
branch itself.

\section{Conclusion}
\label{sec:conclusion}

We have studied whether the dynamical merger branches of massive boson-star
collisions can be inferred from their gravitational waveforms without
supplying the intrinsic source parameters.  Using the existing equal-mass,
head-on numerical-relativity catalogue, we constructed branch-conditioned
reduced-basis surrogates for the real $(2,0)$ and $(4,0)$ modes and used them
as forward models in a parameter-blind waveform search.

On the frozen set of 60 held-out samples, the correct merger branch is
recovered from the $(2,0)$ mode, the $(4,0)$ mode, and their multimode
combination.  Since the dominant $(2,0)$ mode already gives perfect branch
recovery on this test set, the $(4,0)$ result should not be interpreted as an
improvement in classification accuracy.  Rather, it shows that the branch
information is not confined to the dominant mode.  A nearest-template control
gives the same branch assignments, confirming that the separation is already
present in the numerical-relativity waveform manifold rather than being
created by the surrogate.

The parameter-blind analysis also provides source-parameter estimates within
the training-supported catalogue.  The final estimator recovers the correct
branch label for all 60 test samples, gives a mean relative error of $4.30\%$
in $|\phi_c|$, and places $83.3\%$ of the samples within $5\%$ of the true
value.  The self-interaction strength is selected from
$\lambda\in\{10,50,100,300\}$, with an exact recovery rate of $68.3\%$ and
$76.7\%$ recovered within a factor of two.  The remaining parameter errors
are strongly branch dependent and are concentrated mainly in the
$\mathrm{BH}_{\rm pre}$ waveform family.

These results extend the original fixed-parameter proof of concept in two
directions.  First, the merger-route information remains visible in the real
$(4,0)$ higher multipole.  Second, the branch can be inferred without
providing the true $|\phi_c|$, $\lambda$, or merger label during candidate
scoring.  Within the present catalogue, the gravitational waveform therefore
contains information not only about the final outcome, but also about the
dynamical route by which it is reached.

The scope of this conclusion is limited to the present numerical-relativity
data set.  The binaries are equal-mass, head-on, and axisymmetric; only the
real $(2,0)$ and $(4,0)$ modes are included; and the inference is performed
on noiseless processed waveforms.  The value of $\lambda$ is also selected
only from the four simulated catalogue values.  The reported branch accuracy
should therefore not be interpreted as an observational classification
accuracy.

A natural next step is to extend the numerical-relativity catalogue to more
general binary configurations and to include detector response and noise.
This would test whether the waveform-space branch separation survives beyond
the present idealised setting and would help clarify the remaining
source-parameter ambiguity.  Within the current scope, the main conclusion is
that the three dynamical merger routes are encoded sufficiently strongly in
the gravitational radiation to support parameter-blind branch diagnosis on
held-out numerical-relativity waveforms.

\begin{acknowledgments}
The author acknowledges HIAS for access to the ``Quantum Universe
Physical Simulation Platform''. This work used the DiRAC Memory
Intensive service (COSMA8) at Durham University, managed by the
Institute for Computational Cosmology on behalf of the STFC DiRAC HPC
Facility (www.dirac.ac.uk). The DiRAC service at Durham was funded by
BEIS, UKRI and STFC capital funding, Durham University and STFC
operations grants. DiRAC is part of the UKRI Digital Research
Infrastructure. This work was supported by the National Natural Science
Foundation of China under Grant No.~12505066.
\end{acknowledgments}

\clearpage
\appendix

\section{Implementation and reproducibility details}
\label{app:implementation}

This appendix collects the numerical choices needed to reproduce the frozen
benchmark and clarifies the separation between training, validation, and test
data.

\subsection{Frozen data split}
\label{app:frozen_split}

The final catalogue contains 415 selected samples and is divided once into
331 training, 24 validation, and 60 test samples.  The same split is used for
both waveform modes and for all network seeds.  The validation set contains
two samples for each $(\lambda,s)$ cell, while the test set contains five,
so that
\begin{equation}
N_{\rm test}
=
4\times3\times5
=
60 .
\end{equation}
The split is stored as a fixed manifest and is not regenerated between runs.

Five training seeds are used,
\begin{equation}
\{1234,\;2345,\;3456,\;4567,\;5678\}.
\end{equation}

All data-dependent transformations are determined without access to the test
set.  The parameter scaler, training-derived waveform alignment,
branch-specific PCA bases, latent-variable scalers, and the parameter support
used for the blind candidate bank are obtained from the training samples
only.  The validation set is used for checkpoint selection and for choosing
the late-fusion weights.  The latter are selected after grouping validation
samples by the standalone $(4,0)$ branch gate, rather than by their true
branch labels.  The test set is reserved for the final evaluation.

\subsection{Frozen numerical settings}
\label{app:frozen_settings}

The main numerical settings are collected in
Table~\ref{tab:implementation_settings}.  Both waveform modes are represented
on a common grid of 1024 samples with time-grid spacing
$\delta t_{\rm grid}=0.25$.  Parameter-blind inference uses 250 uniformly
spaced values of $|\phi_c|$ within each training-supported
$(\lambda,s)$ interval.

\begin{table}[t]
\centering
\begin{tabular}{lc}
\hline\hline
Setting & Value \\
\hline
Training samples & 331 \\
Validation samples & 24 \\
Test samples & 60 \\
Waveform samples, $N_t$ & 1024 \\
Time-grid spacing, $\delta t_{\rm grid}$ & 0.25 \\
PCA dimension, $K$ & 16 \\
Hidden width & 128 \\
Maximum epochs & 1500 \\
Batch size & 64 \\
Initial learning rate & $10^{-3}$ \\
Weight decay & $10^{-5}$ \\
Early-stopping patience & 180 \\
Maximum alignment shift & 60 \\
Minimum alignment overlap & 0.75 \\
$|\phi_c|$ candidates per $(\lambda,s)$ & 250 \\
Maximum inference shift & 60 \\
Minimum inference overlap & 0.75 \\
Joint-mode weights & $0.5/0.5$ \\
Training seeds & 5 \\
\hline\hline
\end{tabular}
\caption{
Frozen numerical settings used in the final benchmark.  Alignment and
inference shifts are given in the time coordinate of the processed waveforms;
with $\delta t_{\rm grid}=0.25$, a shift of 60 corresponds to 240 grid
points.  In the symmetric $(2,0)+(4,0)$ benchmark, the two modes have equal
weight and share one nuisance time shift.
}
\label{tab:implementation_settings}
\end{table}

Training uses AdamW with a cosine learning-rate schedule and branch-balanced
sampling.  The architecture, reduced-basis dimension, data split, and
inference settings are fixed before evaluation on the held-out test set.

The blind search is restricted to the support of the training catalogue.
For each supported $(\lambda,s)$ pair, the lower and upper values of
$|\phi_c|$ are taken from the corresponding training samples.  The candidate
values of the self-interaction strength are
\begin{equation}
\lambda\in\{10,50,100,300\}.
\end{equation}
The resulting test therefore probes interpolation within the catalogue rather
than extrapolation beyond its sampled parameter range.

\subsection{Reduced-basis representation check}
\label{app:pca_check}

Table~\ref{tab:pca_variance} shows the cumulative variance retained by the
first $K=16$ principal components.  The PCA bases are constructed separately
for each merger branch from training samples only.

In addition to the standalone $(2,0)$ and $(4,0)$ representations, the table
includes the common-frame $(4,0)$ representation used in the symmetric
multimode benchmark.  In that case the $(4,0)$ waveform is represented in the
reference frame associated with the $(2,0)$ mode so that the two channels can
be compared using one common nuisance time shift.  The final asymmetric
parameter estimator instead uses the standalone $(4,0)$ representation.

\begin{table}[t]
\centering
\begin{tabular}{lcc}
\hline\hline
Representation & Branch & Cumulative variance \\
\hline
$(2,0)$
& $\mathrm{BS}_{\rm post}$ & 0.999766 \\
$(2,0)$
& $\mathrm{BH}_{\rm post}$ & 0.999366 \\
$(2,0)$
& $\mathrm{BH}_{\rm pre}$ & 0.999998 \\
\hline
$(4,0)$
& $\mathrm{BS}_{\rm post}$ & 0.992602 \\
$(4,0)$
& $\mathrm{BH}_{\rm post}$ & 0.992127 \\
$(4,0)$
& $\mathrm{BH}_{\rm pre}$ & 0.998870 \\
\hline
$(4,0)$, common frame
& $\mathrm{BS}_{\rm post}$ & 0.990741 \\
$(4,0)$, common frame
& $\mathrm{BH}_{\rm post}$ & 0.996191 \\
$(4,0)$, common frame
& $\mathrm{BH}_{\rm pre}$ & 0.999367 \\
\hline\hline
\end{tabular}
\caption{
Cumulative explained variance retained by the first $K=16$ PCA components.
Each basis is fitted using training samples from the corresponding merger
branch.  The common-frame $(4,0)$ representation is used only for the
symmetric multimode benchmark.
}
\label{tab:pca_variance}
\end{table}

The retained variance exceeds $0.99$ for every branch and representation.
The choice $K=16$ therefore captures the dominant waveform variation in both
modes without requiring a large latent representation.  This dimension is
fixed before the held-out test set is evaluated.

\section{Representative multimode reconstructions}
\label{app:reconstructions}

This appendix gives representative time-domain reconstructions of the two
waveform modes.  The examples are intended as a qualitative check of the
forward surrogate and are not used for model selection or for defining the
inference procedure.

We choose one held-out sample from each merger branch at $\lambda=100$.
Within each branch, the five frozen test samples are ordered by $|\phi_c|$,
and the median sample is selected.  This gives
\begin{equation}
\begin{aligned}
\mathrm{BS}_{\rm post}:&\quad |\phi_c|=0.016,\\
\mathrm{BH}_{\rm post}:&\quad |\phi_c|=0.048,\\
\mathrm{BH}_{\rm pre}:&\quad |\phi_c|=0.150.
\end{aligned}
\end{equation}
The reconstructions use the pre-declared benchmark seed $1234$; no example is
chosen according to its reconstruction error.

\begin{figure*}[t]
\centering
\includegraphics[width=0.96\textwidth]
{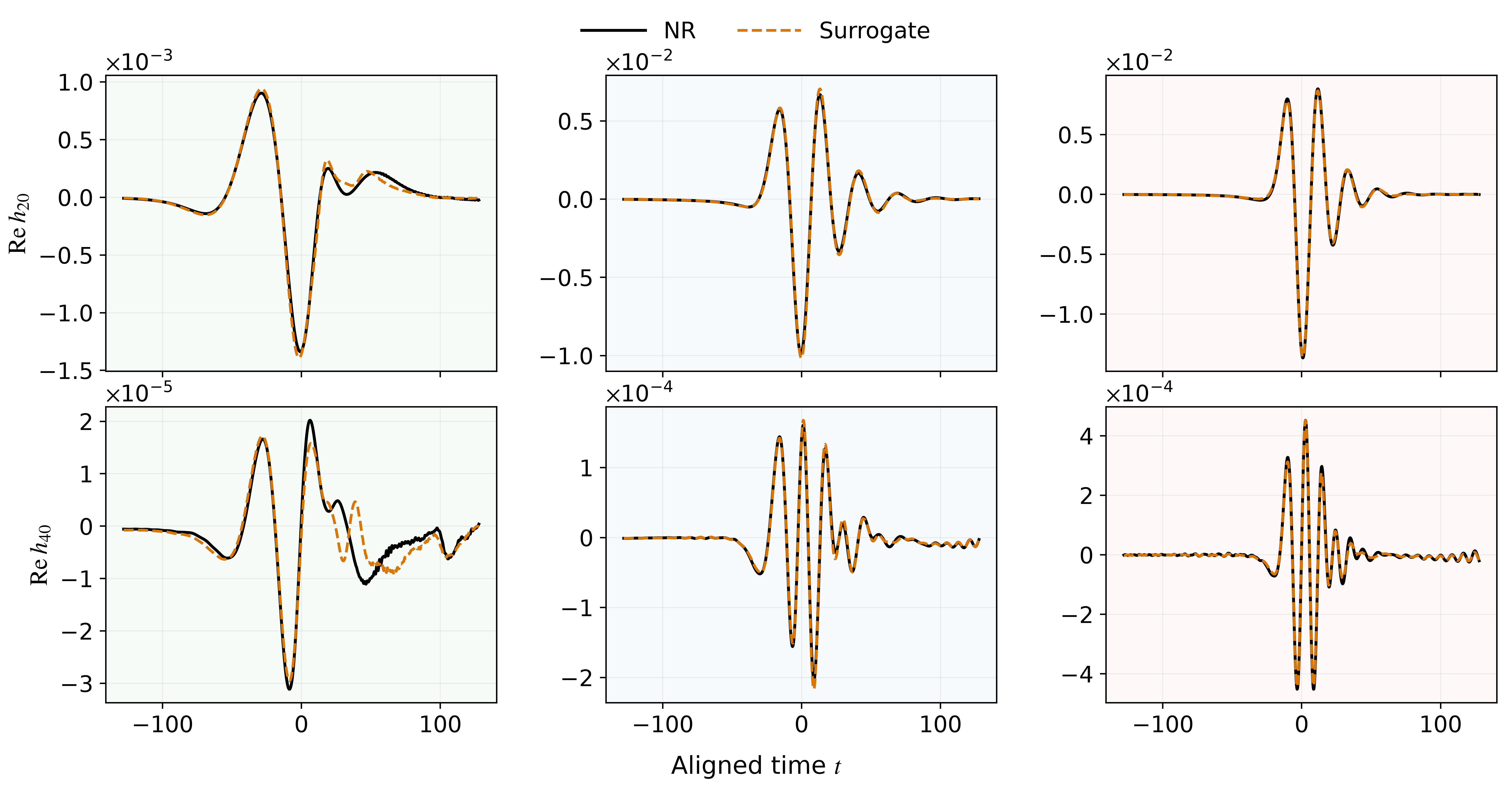}
\caption{
Representative branch-given reconstructions at $\lambda=100$.
From left to right, the columns show
$\mathrm{BS}_{\rm post}$ with $|\phi_c|=0.016$,
$\mathrm{BH}_{\rm post}$ with $|\phi_c|=0.048$, and
$\mathrm{BH}_{\rm pre}$ with $|\phi_c|=0.150$.
The upper row shows $\mathrm{Re}\,h_{20}$ and the lower row
$\mathrm{Re}\,h_{40}$.  Black solid curves are the numerical-relativity
waveforms and orange dashed curves are the surrogate reconstructions.
For each branch, the median-$|\phi_c|$ sample among the five frozen test
samples is shown, using network seed $1234$.  The $(4,0)$ panels use the
common $(2,0)$ reference frame of the symmetric multimode benchmark.
}
\label{fig:representative_multimode_reconstruction}
\end{figure*}

Figure~\ref{fig:representative_multimode_reconstruction} shows the real
$(2,0)$ mode in the upper row and the real $(4,0)$ mode in the lower row.
The $(2,0)$ reconstructions closely follow the dominant burst and subsequent
oscillatory structure in all three branches.  The $(4,0)$ signal is weaker
and more difficult to reproduce in detail, with the largest visible
difference occurring in the late-time part of the
$\mathrm{BS}_{\rm post}$ example.  The principal higher-mode structure is
nevertheless retained in all three cases.

For the lower row, the $(4,0)$ waveform is represented in the common
$(2,0)$ reference frame used in the symmetric multimode benchmark of
Sec.~\ref{subsec:multimode_inference}.  The same training-derived centre time
is therefore used for both modes.  These common-frame reconstructions should
be distinguished from the standalone $(4,0)$ reconstruction statistics
reported in Table~\ref{tab:multimode_branch_results}; the final asymmetric
inference rule also uses the standalone $(4,0)$ representation.

These examples also illustrate why detailed reconstruction accuracy and
branch separability need not coincide.  The $(4,0)$ mode is more difficult
to reconstruct than the dominant mode, yet the standalone $(4,0)$ search
still recovers all 60 branch labels on the frozen test set.  The branch-level
morphology therefore remains distinguishable even when the detailed
time-domain reconstruction is less accurate.

\section{Additional inference checks}
\label{app:controls}

This appendix gives two additional checks on the inference results.  The
first tests whether the merger branches are already separated in waveform
space without neural interpolation.  The second compares the final
$|\phi_c|$ estimate with the standalone $(4,0)$ result.

\subsection{Waveform-space branch controls}
\label{app:template_control}

As a non-neural control, we replace the surrogate candidate waveforms by the
labelled waveforms in the training catalogue.  For each candidate branch, we
take the training waveform with the largest profiled score and then compare
the three branch-wise maxima.  The test waveform is not itself included in
the template set.

Table~\ref{tab:nearest_template_control} summarises the resulting
nearest-template branch diagnosis.

\begin{table}[t]
\centering
\begin{tabular}{lccc}
\hline\hline
Waveform input
& Hybrid accuracy
& Shape-only accuracy
& $\langle\Delta S\rangle$
\\
\hline
$(2,0)$
& $60/60$
& $60/60$
& $0.19346$
\\
$(4,0)$
& $60/60$
& $60/60$
& $0.26511$
\\
$(2,0)+(4,0)$
& $60/60$
& $60/60$
& $0.24685$
\\
\hline\hline
\end{tabular}
\caption{
Nearest-template branch diagnosis on the 60 held-out test samples.  For each
candidate branch, the best-matching waveform is selected from the training
catalogue.  The hybrid and shape-only scores recover every branch label for
the $(2,0)$ mode, the $(4,0)$ mode, and the symmetric multimode comparison.
The final column gives the mean hybrid-score separation between the preferred
branch and its nearest competitor.
}
\label{tab:nearest_template_control}
\end{table}

The branch separation therefore does not depend on neural interpolation.
It is already present when the held-out signals are compared directly with
the training waveform manifold.  The shape-only result gives the same branch
assignments, showing that the separation is not driven solely by the overall
waveform amplitude.  The surrogate provides a smooth branch-conditioned
representation of this waveform-space structure rather than creating the
structure itself.

\subsection{Check on the final $|\phi_c|$ fusion}
\label{app:phic_fusion_check}

The final $|\phi_c|$ estimator combines the independently profiled $(2,0)$
and $(4,0)$ scores using the validation-selected branch-dependent weights of
Sec.~\ref{subsec:final_inference}.  As a diagnostic, we compare this final
estimate with the standalone $(4,0)$ result on the same 60 held-out samples.

We use paired resampling of the test samples to assess the stability of the
difference between the two estimators.  For error measures, the improvement
is defined as
\begin{equation}
\Delta_{\rm imp}
=
E_{40}-E_{\rm fusion},
\end{equation}
so that a positive value favours the final fusion.  For threshold fractions,
the sign convention is reversed accordingly: the fused fraction minus the
standalone $(4,0)$ fraction is reported.

\begin{table*}[t]
\centering
\begin{tabular}{lccc}
\hline\hline
Metric
& Standalone $(4,0)$
& Final fusion
& Median improvement [95\% interval]
\\
\hline
MAE in $|\phi_c|$
& $0.00600$
& $0.005735$
& $0.000268\;
[-7.49\times10^{-5},\,5.69\times10^{-4}]$
\\
Mean relative error
& $6.54\%$
& $4.30\%$
& $2.20$ percentage points
$\;[0.97,\,3.85]$
\\
Within $5\%$
& $66.7\%$
& $83.3\%$
& $16.67$ percentage points
$\;[8.33,\,26.67]$
\\
Within $10\%$
& $78.3\%$
& $86.7\%$
& $8.33$ percentage points
$\;[0,\,16.67]$
\\
\hline\hline
\end{tabular}
\caption{
Comparison of the final branch-conditioned $|\phi_c|$ fusion with the
standalone $(4,0)$ estimator.  The intervals are obtained by paired
resampling of the 60 held-out samples.  Positive improvements favour the
final estimator.  The MAE interval crosses zero, whereas the intervals for
the mean relative error and the fraction recovered within $5\%$ remain
positive.
}
\label{tab:phic_bootstrap_control}
\end{table*}

The final fusion lowers the mean relative error and increases the fraction of
samples recovered within $5\%$ of the true $|\phi_c|$.  These differences
remain positive over the quoted 95\% resampling intervals.  The evidence is
weaker for the within-$10\%$ fraction, whose interval reaches zero.  The MAE
comparison is also less conclusive: although the point estimate is smaller
for the fused estimator, the corresponding interval crosses zero.

This check therefore supports the more limited statement used in the main
text: the multimode late fusion improves several measures of
$|\phi_c|$ recovery, but it is not uniformly superior to the standalone
$(4,0)$ estimator under every metric.  The $\lambda$ result is unaffected by
this comparison because the final $\lambda$ estimate is, by construction,
the standalone $(4,0)$ estimate.

\clearpage
\bibliography{BS}

@article{Palenzuela:2017kcg,
  author = "Palenzuela, C. and Pani, P. and Bezares, M. and Cardoso, V. and Lehner, L. and Liebling, S.",
  title = "{Gravitational Wave Signatures of Highly Compact Boson Star Binaries}",
  journal = "Phys. Rev. D",
  volume = "96",
  year = "2017",
  number = "10",
  pages = "104058",
  doi = "10.1103/PhysRevD.96.104058",
  eprint = "arXiv:1710.09432 [gr-qc]"}

@article{Andrade:2021rbd,
    author = "Andrade, Tomas and others",
    title = "{GRChombo: An adaptable numerical relativity code for fundamental physics}",
    eprint = "2201.03458",
    archivePrefix = "arXiv",
    primaryClass = "gr-qc",
    doi = "10.21105/joss.03703",
    journal = "J. Open Source Softw.",
    volume = "6",
    number = "68",
    pages = "3703",
    year = "2021"
}

@article{Kaup:1968zz,
    author = "Kaup, David J.",
    title = "{Klein-Gordon Geon}",
    doi = "10.1103/PhysRev.172.1331",
    journal = "Phys. Rev.",
    volume = "172",
    pages = "1331--1342",
    year = "1968"
}

@article{Helfer:2021brt,
    author = "Helfer, Thomas and Sperhake, Ulrich and Croft, Robin and Radia, Miren and Ge, Bo-Xuan and Lim, Eugene A.",
    title = "{Malaise and remedy of binary boson-star initial data}",
    eprint = "2108.11995",
    archivePrefix = "arXiv",
    primaryClass = "gr-qc",
    doi = "10.1088/1361-6382/ac53b7",
    journal = "Class. Quant. Grav.",
    volume = "39",
    number = "7",
    pages = "074001",
    year = "2022"
}

@article{Yoshida:1997qf,
    author = "Yoshida, Shijun and Eriguchi, Yoshiharu",
    title = "{Rotating boson stars in general relativity}",
    reportNumber = "PRINT-97-136 (TOKYO)",
    doi = "10.1103/PhysRevD.56.762",
    journal = "Phys. Rev. D",
    volume = "56",
    pages = "762--771",
    year = "1997"
}

@article{Siemonsen:2020hcg,
    author = "Siemonsen, Nils and East, William E.",
    title = "{Stability of rotating scalar boson stars with nonlinear interactions}",
    eprint = "2011.08247",
    archivePrefix = "arXiv",
    primaryClass = "gr-qc",
    doi = "10.1103/PhysRevD.103.044022",
    journal = "Phys. Rev. D",
    volume = "103",
    number = "4",
    pages = "044022",
    year = "2021"
}

@article{Sanchis-Gual:2019ljs,
    author = "Sanchis-Gual, N. and Di Giovanni, F. and Zilhão, M. and Herdeiro, C. and Cerdá-Durán, P. and Font, J.A. and Radu, E.",
    title = "{Nonlinear Dynamics of Spinning Bosonic Stars: Formation and Stability}",
    eprint = "1907.12565",
    archivePrefix = "arXiv",
    primaryClass = "gr-qc",
    doi = "10.1103/PhysRevLett.123.221101",
    journal = "Phys. Rev. Lett.",
    volume = "123",
    number = "22",
    pages = "221101",
    year = "2019"
}

@article{Cook:2016soy,
    author = "Cook, William G. and Figueras, Pau and Kunesch, Markus and Sperhake, Ulrich and Tunyasuvunakool, Saran",
    editor = "Herdeiro, Carlos A. R. and Berti, Emanuele and Cardoso, Vitor and Crispino, Luis C. B. and Gualtieri, Leonardo and Sperhake, Ulrich",
    title = "{Dimensional reduction in numerical relativity: Modified cartoon formalism and regularization}",
    eprint = "1603.00362",
    archivePrefix = "arXiv",
    primaryClass = "gr-qc",
    doi = "10.1142/S0218271816410133",
    journal = "Int. J. Mod. Phys. D",
    volume = "25",
    number = "09",
    pages = "1641013",
    year = "2016"
}

@article{Abbott:2016blz,
  author = "Abbott, B. P. and others",
  title = "{Observation of Gravitational Waves from a Binary Black Hole Merger}",
  journal = "Phys. Rev. Lett.",
  volume = "116",
  year = "2016",
  number = "6",
  pages = "061102",
  doi = "10.1103/PhysRevLett.116.061102",
  eprint = "1602.03837",
  archivePrefix = "arXiv",
  primaryClass = "gr-qc",
  reportNumber = "LIGO-P150914"}

@article{LIGOScientific:2018mvr,
  author = "Abbott, B. P. and others",
  collaboration = "LIGO Scientific, Virgo",
  title = "{GWTC-1: A Gravitational-Wave Transient Catalog of Compact Binary Mergers Observed by LIGO and Virgo during the First and Second Observing Runs}",
  eprint = "1811.12907",
  archivePrefix = "arXiv",
  primaryClass = "astro-ph.HE",
  reportNumber = "LIGO-P1800307",
  doi = "10.1103/PhysRevX.9.031040",
  journal = "Phys. Rev. X",
  volume = "9",
  number = "3",
  pages = "031040",
  year = "2019"}

@article{LIGOScientific:2020ibl,
  author = "Abbott, R. and others",
  collaboration = "LIGO Scientific, Virgo",
  title = "{GWTC-2: Compact Binary Coalescences Observed by LIGO and Virgo During the First Half of the Third Observing Run}",
  eprint = "2010.14527",
  archivePrefix = "arXiv",
  primaryClass = "gr-qc",
  reportNumber = "P2000061",
  doi = "10.1103/PhysRevX.11.021053",
  journal = "Phys. Rev. X",
  volume = "11",
  pages = "021053",
  year = "2021"}

@article{LIGOScientific:2021djp,
  author = "Abbott, R. and others",
  collaboration = "LIGO Scientific, VIRGO, KAGRA",
  title = "{GWTC-3: Compact Binary Coalescences Observed by LIGO and Virgo During the Second Part of the Third Observing Run}",
  eprint = "2111.03606",
  archivePrefix = "arXiv",
  primaryClass = "gr-qc",
  reportNumber = "LIGO-P2000318",
  month = "11",
  year = "2021"}

@article{Cook:2016qnt,
    author = "Cook, William G. and Sperhake, Ulrich",
    title = "{Extraction of gravitational-wave energy in higher dimensional numerical relativity using the Weyl tensor}",
    eprint = "1609.01292",
    archivePrefix = "arXiv",
    primaryClass = "gr-qc",
    doi = "10.1088/1361-6382/aa5294",
    journal = "Class. Quant. Grav.",
    volume = "34",
    number = "3",
    pages = "035010",
    year = "2017"
}

@article{Seidel:1993zk,
    author = "Seidel, Edward and Suen, Wai-Mo",
    title = "{Formation of solitonic stars through gravitational cooling}",
    eprint = "gr-qc/9309015",
    archivePrefix = "arXiv",
    reportNumber = "PRINT-93-0724 (NCSA,URBANA)",
    doi = "10.1103/PhysRevLett.72.2516",
    journal = "Phys. Rev. Lett.",
    volume = "72",
    pages = "2516--2519",
    year = "1994"
}

@article{Palenzuela:2006wp,
    author = "Palenzuela, C. and Olabarrieta, I. and Lehner, L. and Liebling, Steven L.",
    title = "{Head-on collisions of boson stars}",
    eprint = "gr-qc/0612067",
    archivePrefix = "arXiv",
    doi = "10.1103/PhysRevD.75.064005",
    journal = "Phys. Rev. D",
    volume = "75",
    pages = "064005",
    year = "2007"
}

@article{Palenzuela:2007dm,
    author = "Palenzuela, C. and Lehner, L. and Liebling, Steven L.",
    title = "{Orbital Dynamics of Binary Boson Star Systems}",
    eprint = "0706.2435",
    archivePrefix = "arXiv",
    primaryClass = "gr-qc",
    doi = "10.1103/PhysRevD.77.044036",
    journal = "Phys. Rev. D",
    volume = "77",
    pages = "044036",
    year = "2008"
}

@article{Seidel:1990jh,
    author = "Seidel, Edward and Suen, Wai-Mo",
    title = "{Dynamical Evolution of Boson Stars. 1. Perturbing the Ground State}",
    reportNumber = "PRINT-90-0131 (WASH.U.,ST.LOUIS)",
    doi = "10.1103/PhysRevD.42.384",
    journal = "Phys. Rev. D",
    volume = "42",
    pages = "384--403",
    year = "1990"
}

@article{Balakrishna:1997ej,
    author = "Balakrishna, Jayashree and Seidel, Edward and Suen, Wai-Mo",
    title = "{Dynamical evolution of boson stars. 2. Excited states and selfinteracting fields}",
    eprint = "gr-qc/9712064",
    archivePrefix = "arXiv",
    doi = "10.1103/PhysRevD.58.104004",
    journal = "Phys. Rev. D",
    volume = "58",
    pages = "104004",
    year = "1998"
}

@article{Siemonsen:2023hko,
    author = "Siemonsen, Nils and East, William E.",
    title = "{Binary boson stars: Merger dynamics and formation of rotating remnant stars}",
    eprint = "2302.06627",
    archivePrefix = "arXiv",
    primaryClass = "gr-qc",
    doi = "10.1103/PhysRevD.107.124018",
    journal = "Phys. Rev. D",
    volume = "107",
    number = "12",
    pages = "124018",
    year = "2023"
}

@article{Evstafyeva:2022bpr,
    author = "Evstafyeva, Tamara and Sperhake, Ulrich and Helfer, Thomas and Croft, Robin and Radia, Miren and Ge, Bo-Xuan and Lim, Eugene A.",
    title = "{Unequal-mass boson-star binaries: initial data and merger dynamics}",
    eprint = "2212.08023",
    archivePrefix = "arXiv",
    primaryClass = "gr-qc",
    doi = "10.1088/1361-6382/acc2a8",
    journal = "Class. Quant. Grav.",
    volume = "40",
    number = "8",
    pages = "085009",
    year = "2023"
}

@article{Siemonsen:2023age,
    author = "Siemonsen, Nils and East, William E.",
    title = "{Generic initial data for binary boson stars}",
    eprint = "2306.17265",
    archivePrefix = "arXiv",
    primaryClass = "gr-qc",
    doi = "10.1103/PhysRevD.108.124015",
    journal = "Phys. Rev. D",
    volume = "108",
    number = "12",
    pages = "124015",
    year = "2023"
}

@article{Bezares:2022obu,
    author = "Bezares, Miguel and Bo\v{s}kovi\'c, Mateja and Liebling, Steven and Palenzuela, Carlos and Pani, Paolo and Barausse, Enrico",
    title = "{Gravitational waves and kicks from the merger of unequal mass, highly compact boson stars}",
    eprint = "2201.06113",
    archivePrefix = "arXiv",
    primaryClass = "gr-qc",
    doi = "10.1103/PhysRevD.105.064067",
    journal = "Phys. Rev. D",
    volume = "105",
    number = "6",
    pages = "064067",
    year = "2022"
}

@article{Ruffini:1969qy,
    author = "Ruffini, Remo and Bonazzola, Silvano",
    title = "{Systems of selfgravitating particles in general relativity and the concept of an equation of state}",
    doi = "10.1103/PhysRev.187.1767",
    journal = "Phys. Rev.",
    volume = "187",
    pages = "1767--1783",
    year = "1969"
}

@article{Colpi:1986ye,
    author = "Colpi, M. and Shapiro, S. L. and Wasserman, I.",
    title = "{Boson Stars: Gravitational Equilibria of Selfinteracting Scalar Fields}",
    doi = "10.1103/PhysRevLett.57.2485",
    journal = "Phys. Rev. Lett.",
    volume = "57",
    pages = "2485--2488",
    year = "1986"
}

@article{Schunck:1999zu,
    author = "Schunck, Franz E. and Torres, Diego F.",
    title = "{Boson stars with generic selfinteractions}",
    eprint = "gr-qc/9911038",
    archivePrefix = "arXiv",
    doi = "10.1142/S0218271800000608",
    journal = "Int. J. Mod. Phys. D",
    volume = "9",
    pages = "601--618",
    year = "2000"
}

@article{Hartmann:2012da,
    author = "Hartmann, Betti and Kleihaus, Burkhard and Kunz, Jutta and Schaffer, Isabell",
    title = "{Compact Boson Stars}",
    eprint = "1205.0899",
    archivePrefix = "arXiv",
    primaryClass = "gr-qc",
    doi = "10.1016/j.physletb.2012.06.067",
    journal = "Phys. Lett. B",
    volume = "714",
    pages = "120--126",
    year = "2012"
}

@article{Kobayashi:1994qi,
    author = "Kobayashi, Y. and Kasai, M. and Futamase, T.",
    title = "{Does a boson star rotate?}",
    doi = "10.1103/PhysRevD.50.7721",
    journal = "Phys. Rev. D",
    volume = "50",
    pages = "7721--7724",
    year = "1994"
}

@article{Ryan:1996nk,
    author = "Ryan, Fintan D.",
    title = "{Spinning boson stars with large selfinteraction}",
    reportNumber = "PRINT-97-067 (CAL-TECH)",
    doi = "10.1103/PhysRevD.55.6081",
    journal = "Phys. Rev. D",
    volume = "55",
    pages = "6081--6091",
    year = "1997"
}

@article{Schunck:1996he,
    author = "Schunck, France E. and Mielke, Eckehard W.",
    title = "{Rotating boson star as an effective mass torus in general relativity}",
    doi = "10.1016/S0375-9601(98)00778-6",
    journal = "Phys. Lett. A",
    volume = "249",
    pages = "389--394",
    year = "1998"
}

@article{Schunck:2003kk,
    author = "Schunck, Franz E. and Mielke, Eckehard W.",
    title = "{General relativistic boson stars}",
    eprint = "0801.0307",
    archivePrefix = "arXiv",
    primaryClass = "astro-ph",
    doi = "10.1088/0264-9381/20/20/201",
    journal = "Class. Quant. Grav.",
    volume = "20",
    pages = "R301--R356",
    year = "2003"
}

@article{Balakrishna:2006ru,
    author = "Balakrishna, Jayashree and Bondarescu, Ruxandra and Daues, Gregory and Siddhartha Guzman, F. and Seidel, Edward",
    title = "{Evolution of 3-D boson stars with waveform extraction}",
    eprint = "gr-qc/0602078",
    archivePrefix = "arXiv",
    doi = "10.1088/0264-9381/23/7/024",
    journal = "Class. Quant. Grav.",
    volume = "23",
    pages = "2631--2652",
    year = "2006"
}

@article{Balakrishna:2007mr,
    author = "Balakrishna, Jayashree and Bondarescu, Ruxandra and Daues, Gregory and Bondarescu, Mihai",
    title = "{Numerical Simulations of Oscillating Soliton Stars: Excited States in Spherical Symmetry and Ground State Evolutions in 3D}",
    eprint = "0710.4131",
    archivePrefix = "arXiv",
    primaryClass = "gr-qc",
    doi = "10.1103/PhysRevD.77.024028",
    journal = "Phys. Rev. D",
    volume = "77",
    pages = "024028",
    year = "2008"
}

@article{Schunck:1999pm,
    author = "Schunck, F. E. and Mielke, E. W.",
    title = "{Boson stars: Rotation, formation, and evolution}",
    doi = "10.1023/A:1026673918588",
    journal = "Gen. Rel. Grav.",
    volume = "31",
    pages = "787",
    year = "1999"
}

@article{Sanchis-Gual:2020mzb,
    author = "Sanchis-Gual, Nicolas and Zilh{\~a}o, Miguel and Herdeiro, Carlos and Di Giovanni, Fabrizio and Font, Jos{\'e} A. and Radu, Eugen",
    title = "{Synchronized gravitational atoms from mergers of bosonic stars}",
    eprint = "2007.11584",
    archivePrefix = "arXiv",
    primaryClass = "gr-qc",
    doi = "10.1103/PhysRevD.102.101504",
    journal = "Phys. Rev. D",
    volume = "102",
    number = "10",
    pages = "101504",
    year = "2020"
}

@article{Croft:2022bxq,
    author = "Croft, Robin and Helfer, Thomas and Ge, Bo-Xuan and Radia, Miren and Evstafyeva, Tamara and Lim, Eugene A. and Sperhake, Ulrich and Clough, Katy",
    title = "{The gravitational afterglow of boson stars}",
    eprint = "2207.05690",
    archivePrefix = "arXiv",
    primaryClass = "gr-qc",
    reportNumber = "KCL-TH-2021-82",
    doi = "10.1088/1361-6382/acace4",
    journal = "Class. Quant. Grav.",
    volume = "40",
    number = "6",
    pages = "065001",
    year = "2023"
}

@article{Sanchis-Gual:2022zsr,
    author = "Sanchis-Gual, Nicolas and Zilh{\~a}o, Miguel and Cardoso, Vitor",
    title = "{Electromagnetic emission from axionic boson star collisions}",
    eprint = "2207.05494",
    archivePrefix = "arXiv",
    primaryClass = "gr-qc",
    doi = "10.1103/PhysRevD.106.064034",
    journal = "Phys. Rev. D",
    volume = "106",
    number = "6",
    pages = "064034",
    year = "2022"
}

@phdthesis{Ge:2024fum,
    author = "Ge, Bo-Xuan",
    title = "{Gravitational Waves in Boson Star Mergers}",
    school = "King's College London",
    month = "10",
    year = "2024"
}

@article{Ge:2024itl,
    author = "Ge, Bo-Xuan and Lim, Eugene A. and Sperhake, Ulrich and Evstafyeva, Tamara and Cors, Daniela and de Jong, Eloy and Croft, Robin and Helfer, Thomas",
    title = "{Dynamics and gravitational radiation of stable and unstable boson-star mergers}",
    eprint = "2410.23839",
    archivePrefix = "arXiv",
    primaryClass = "gr-qc",
    reportNumber = "KCL-TH-2024-59",
    doi = "10.1103/2dhs-phl4",
    journal = "Phys. Rev. D",
    volume = "112",
    number = "12",
    pages = "124080",
    year = "2025"
}

@article{Evstafyeva:2025mvx,
    author = "Evstafyeva, Tamara and Siemonsen, Nils and East, William E.",
    title = "{Assessing the stability of ultracompact spinning boson stars with nonlinear evolutions}",
    eprint = "2508.11527",
    archivePrefix = "arXiv",
    primaryClass = "gr-qc",
    doi = "10.1103/2vz3-s39r",
    journal = "Phys. Rev. D",
    volume = "113",
    number = "4",
    pages = "044024",
    year = "2026"
}

@article{Marks:2025jpt,
    author = "Marks, Gareth Arturo and Staelens, Seppe J. and Evstafyeva, Tamara and Sperhake, Ulrich",
    title = "{Long-Term Stable Nonlinear Evolutions of Ultracompact Black-Hole Mimickers}",
    eprint = "2504.17775",
    archivePrefix = "arXiv",
    primaryClass = "gr-qc",
    doi = "10.1103/lk48-7r2f",
    journal = "Phys. Rev. Lett.",
    volume = "135",
    number = "13",
    pages = "131402",
    year = "2025"
}

@article{Evstafyeva:2024qvp,
    author = "Evstafyeva, Tamara and Sperhake, Ulrich and Romero-Shaw, Isobel M. and Agathos, Michalis",
    title = "{Gravitational-Wave Data Analysis with High-Precision Numerical Relativity Simulations of Boson Star Mergers}",
    eprint = "2406.02715",
    archivePrefix = "arXiv",
    primaryClass = "gr-qc",
    doi = "10.1103/PhysRevLett.133.131401",
    journal = "Phys. Rev. Lett.",
    volume = "133",
    number = "13",
    pages = "131401",
    year = "2024"
}

@article{Evstafyeva:2023kfg,
    author = "Evstafyeva, Tamara and Rosca-Mead, Roxana and Sperhake, Ulrich and Brugmann, Bernd",
    title = "{Boson stars in massless and massive scalar-tensor gravity}",
    eprint = "2310.05200",
    archivePrefix = "arXiv",
    primaryClass = "gr-qc",
    doi = "10.1103/PhysRevD.108.104064",
    journal = "Phys. Rev. D",
    volume = "108",
    number = "10",
    pages = "104064",
    year = "2023"
}

@article{Marks:2025xxv,
    author = "Marks, Gareth Arturo and Zaif, Abdullah Al",
    title = "{Boson stars in $D \unicode{x2A7E} 4$ dimensions: stability, oscillation frequencies, and dynamical evolutions}",
    eprint = "2510.13988",
    archivePrefix = "arXiv",
    primaryClass = "gr-qc",
    doi = "10.1088/1361-6382/ae60bc",
    journal = "Class. Quant. Grav.",
    volume = "43",
    number = "8",
    pages = "085014",
    year = "2026"
}

@article{Marks:2025jit,
    author = "Marks, Gareth Arturo",
    title = "{Perturbations of Solitonic Boson Stars: Nonlinear Radial Stability and Binding Energy}",
    eprint = "2508.11757",
    archivePrefix = "arXiv",
    primaryClass = "gr-qc",
    doi = "10.1088/1742-6596/3177/1/012047",
    journal = "J. Phys. Conf. Ser.",
    volume = "3177",
    number = "1",
    pages = "012047",
    year = "2026"
}

@article{Ma:2024olw,
    author = "Ma, Tian-Xiang and Fang, Tie-Feng and Wang, Yong-Qiang",
    title = "{Boson stars and their frozen states in an infinite tower of higher-derivative gravity}",
    eprint = "2406.08813",
    archivePrefix = "arXiv",
    primaryClass = "gr-qc",
    doi = "10.1140/epjc/s10052-025-14252-4",
    journal = "Eur. Phys. J. C",
    volume = "85",
    number = "5",
    pages = "542",
    year = "2025"
}

@article{Ding:2023syj,
    author = "Ding, Peng-Bo and Ma, Tian-Xiang and Fang, Tie-Feng and Wang, Yong-Qiang",
    title = "{Study of boson stars with wormhole}",
    eprint = "2305.19819",
    archivePrefix = "arXiv",
    primaryClass = "gr-qc",
    doi = "10.1007/JHEP04(2024)033",
    journal = "JHEP",
    volume = "04",
    pages = "033",
    year = "2024"
}

@article{Liang:2022mjo,
    author = "Liang, Chen and Ren, Ji-Rong and Sun, Shi-Xian and Wang, Yong-Qiang",
    title = "{Dirac-boson stars}",
    eprint = "2207.11147",
    archivePrefix = "arXiv",
    primaryClass = "gr-qc",
    doi = "10.1007/JHEP02(2023)249",
    journal = "JHEP",
    volume = "02",
    pages = "249",
    year = "2023"
}

@article{deSa:2025nsx,
    author = "de S{\'a}, Pedro L. Brito and Lima, Haroldo C. D. and Herdeiro, Carlos A. R. and Crispino, Lu{\'\i}s C. B.",
    title = "{Spherical Einstein-Friedberg-Lee-Sirlin boson stars: Self-interacting solutions and their astrophysical appearance}",
    eprint = "2511.19206",
    archivePrefix = "arXiv",
    primaryClass = "gr-qc",
    doi = "10.1103/hs1t-8plf",
    journal = "Phys. Rev. D",
    volume = "113",
    number = "4",
    pages = "044037",
    year = "2026"
}

@article{Brito:2025rld,
    author = "Brito, Marco and Herdeiro, Carlos and Radu, Eugen and Sanchis-Gual, Nicolas and Zilh{\~a}o, Miguel",
    title = "{Stability and collisions of excited spherical boson stars: Glimpses of chains and rings}",
    eprint = "2506.06442",
    archivePrefix = "arXiv",
    primaryClass = "gr-qc",
    doi = "10.1103/8bwz-wqhp",
    journal = "Phys. Rev. D",
    volume = "113",
    number = "2",
    pages = "024008",
    year = "2026"
}

@article{Herdeiro:2025lwf,
    author = "Herdeiro, Carlos A. R. and Radu, Eugen",
    title = "{Gregory-Laflamme-type instability of boson strings and related phases in D = 5 Kaluza-Klein theory}",
    eprint = "2503.15069",
    archivePrefix = "arXiv",
    primaryClass = "gr-qc",
    doi = "10.1007/JHEP08(2025)049",
    journal = "JHEP",
    volume = "08",
    pages = "049",
    year = "2025"
}

@article{Herdeiro:2024myz,
    author = "Herdeiro, Carlos and Huang, Hyat and Kunz, Jutta and Radu, Eugen",
    title = "{Einstein-(complex)-Maxwell static boson stars in AdS}",
    eprint = "2405.10671",
    archivePrefix = "arXiv",
    primaryClass = "gr-qc",
    doi = "10.1016/j.physletb.2024.138939",
    journal = "Phys. Lett. B",
    volume = "856",
    pages = "138939",
    year = "2024"
}

@article{Ildefonso:2023qty,
    author = "Ildefonso, Pedro and Zilh{\~a}o, Miguel and Herdeiro, Carlos and Radu, Eugen and Santos, Nuno M.",
    title = "{Self-interacting dipolar boson stars and their dynamics}",
    eprint = "2307.00044",
    archivePrefix = "arXiv",
    primaryClass = "gr-qc",
    doi = "10.1103/PhysRevD.108.064011",
    journal = "Phys. Rev. D",
    volume = "108",
    number = "6",
    pages = "064011",
    year = "2023"
}

@article{Brito:2023fwr,
    author = "Brito, Marco and Herdeiro, Carlos and Radu, Eugen and Sanchis-Gual, Nicolas and Zilh{\~a}o, Miguel",
    title = "{Stability and physical properties of spherical excited scalar boson stars}",
    eprint = "2302.08900",
    archivePrefix = "arXiv",
    primaryClass = "gr-qc",
    doi = "10.1103/PhysRevD.107.084022",
    journal = "Phys. Rev. D",
    volume = "107",
    number = "8",
    pages = "084022",
    year = "2023"
}

@article{Jaramillo:2022zwg,
    author = "Jaramillo, V{\'\i}ctor and Sanchis-Gual, Nicolas and Barranco, Juan and Bernal, Argelia and Degollado, Juan Carlos and Herdeiro, Carlos and Megevand, Miguel and N{\'u}{\~n}ez, Dar{\'\i}o",
    title = "{Head-on collisions of {\ensuremath{\ell}}-boson stars}",
    eprint = "2202.00696",
    archivePrefix = "arXiv",
    primaryClass = "gr-qc",
    doi = "10.1103/PhysRevD.105.104057",
    journal = "Phys. Rev. D",
    volume = "105",
    number = "10",
    pages = "104057",
    year = "2022"
}

@misc{Damour:2025oys,
    author = "Damour, Thibault and Jain, Tamanna and Sperhake, Ulrich",
    title = "{Gravitational scattering of solitonic boson stars: Analytics vs Numerics}",
    eprint = "2512.00945",
    archivePrefix = "arXiv",
    primaryClass = "gr-qc",
    month = "11",
    year = "2025"
}

@article{Zhang:2023qxf,
    author = "Zhang, Yu-Peng and Sun, Shi-Xian and Wang, Yong-Qiang and Wei, Shao-Wen and Laguna, Pablo and Liu, Yu-Xiao",
    title = "{Fate of initially bound timelike geodesics in spherical boson stars}",
    eprint = "2310.01178",
    archivePrefix = "arXiv",
    primaryClass = "gr-qc",
    doi = "10.1103/PhysRevResearch.6.033187",
    journal = "Phys. Rev. Res.",
    volume = "6",
    number = "3",
    pages = "033187",
    year = "2024"
}

@misc{Zhang:2025xnl,
    author        = "Zhang, Yu-Peng and Wei, Shao-Wen and Liu, Yu-Xiao",
    title         = "{Emerging black hole shadow from collapsing boson star}",
    eprint        = "2503.14159",
    archivePrefix = "arXiv",
    primaryClass  = "gr-qc",
    year          = "2025",
    month         = mar
}

@article{Ge:2025btw,
    author = "Ge, Bo-Xuan",
    title = "{Massive boson stars: Stability and GW emission in head-on mergers}",
    eprint = "2512.15242",
    archivePrefix = "arXiv",
    primaryClass = "gr-qc",
    doi = "10.1103/y3t4-5t1q",
    journal = "Phys. Rev. D",
    volume = "113",
    number = "10",
    pages = "104065",
    year = "2026"
}

@misc{Ning:2026qxs,
    author = "Ning, Zhuan",
    title = "{Boson star-black hole binaries: initial data and head-on collisions}",
    eprint = "2604.15240",
    archivePrefix = "arXiv",
    primaryClass = "gr-qc",
    month = "4",
    year = "2026"
}

@article{Ge:2026wzh,
    author = "Ge, Bo-Xuan",
    title = "{Timing-Window Mechanism for Chain-Like Transients in Collisions of Radially Excited Boson Stars}",
    eprint = "2605.19572",
    archivePrefix = "arXiv",
    primaryClass = "gr-qc",
    month = "5",
    year = "2026"
}

@article{Chua:2018woh,
    author = "Chua, Alvin J. K. and Galley, Chad R. and Vallisneri, Michele",
    title = "{Reduced-order modeling with artificial neurons for gravitational-wave inference}",
    eprint = "1811.05491",
    archivePrefix = "arXiv",
    primaryClass = "astro-ph.IM",
    doi = "10.1103/PhysRevLett.122.211101",
    journal = "Phys. Rev. Lett.",
    volume = "122",
    number = "21",
    pages = "211101",
    year = "2019"
}

@article{Cuoco:2024cdk,
    author = "Cuoco, Elena and Cavagli{\`a}, Marco and Heng, Ik Siong and Keitel, David and Messenger, Christopher",
    title = "{Applications of machine learning in gravitational-wave research with current interferometric detectors}",
    eprint = "2412.15046",
    archivePrefix = "arXiv",
    primaryClass = "gr-qc",
    reportNumber = "LIGO-P2400077",
    doi = "10.1007/s41114-024-00055-8",
    journal = "Living Rev. Rel.",
    volume = "28",
    number = "1",
    pages = "2",
    year = "2025"
}

@article{Ge:2026fki,
    author = "Ge, Bo-Xuan",
    title = "{Relative-Phase Control of Common-Bridge Survival in Collisions of Radially Excited Boson Stars}",
    eprint = "2609.01134",
    archivePrefix = "arXiv",
    primaryClass = "gr-qc",
    month = "9",
    year = "2026"
}

\end{document}